# The Subspace Voyager: Exploring High-Dimensional Data along a Continuum of Salient 3D Subspace

Bing Wang and Klaus Mueller, *Senior Member, IEEE*

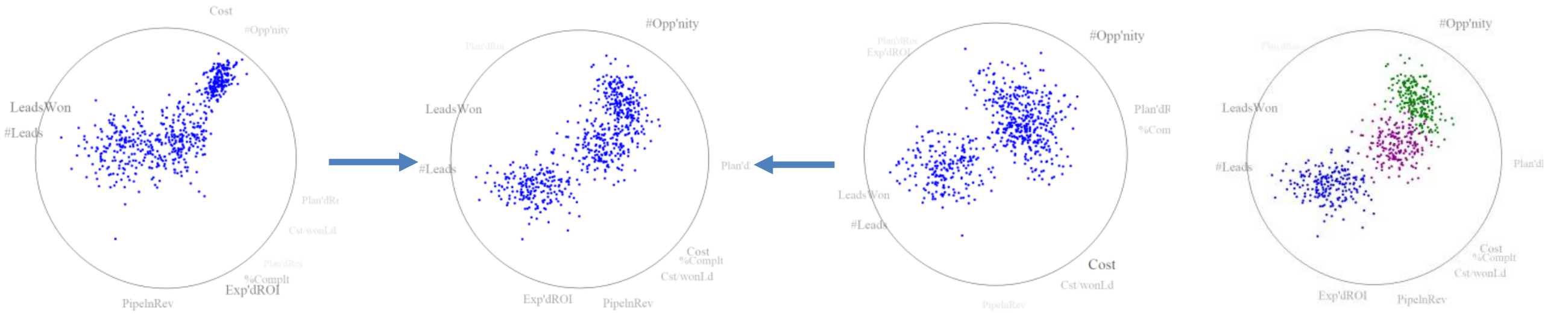

(a) (b) (c) (d)

Panel (a) and (c) are two projective views onto a 10-dimesional sales pipeline dataset with 900 points. The labels at the circle boundary indicate the data attributes and their axis directions in that view. The strength of the label fonts indicates how well the attributes are expressed in this view. Panel (b) shows a view generated by using our system's trackball interface to generate new projective views between view (a) and (c). The motion parallax clarified that there were not two but three clusters. Panel (d) shows the three clusters in different colors.

**Abstract**— Analyzing high-dimensional data and finding hidden patterns is a difficult problem and has attracted numerous research efforts. Automated methods can be useful to some extent but bringing the data analyst into the loop via interactive visual tools can help the discovery process tremendously. An inherent problem in this effort is that humans lack the mental capacity to truly understand spaces exceeding three spatial dimensions. To keep within this limitation, we describe a framework that decomposes a high-dimensional data space into a continuum of generalized 3D subspaces. Analysts can then explore these 3D subspaces individually via the familiar trackball interface while using additional facilities to smoothly transition to adjacent subspaces for expanded space comprehension. Since the number of such subspaces suffers from combinatorial explosion, we provide a set of data-driven subspace selection and navigation tools which can guide users to interesting subspaces and views. A subspace trail map allows users to manage the explored subspaces, keep their bearings, and return to interesting subspaces and views. Both trackball and trail map are each embedded into a word cloud of attribute labels which aid in navigation. We demonstrate our system via several use cases in a diverse set of application areas – cluster analysis and refinement, information discovery, and supervised training of classifiers. We also report on a user study that evaluates the usability of the various interactions our system provides.

## 1. Introduction

DATA with many attributes have become commonplace in a wide range of domains, such as science, business, medicine, etc. In these data, the most interesting relations are often multivariate, and gaining proper tools to recognize these relationships reliably is still an active area of research. While automated analysis can be useful in finding some of the high-dimensional patterns, adding the human into the loop can break ties and help discern patterns in confounding and noisy data settings that benefit from the intricate reasoning faculties of human domain experts. However, we are still far off from having effective visual tools for high-D data analytics that make the best use of the inborn capabilities of the human visual system and at the same time also observe its limitations.

High-D space is generally confusing to most people since humans do not possess the innate neural network to recognize and reason with high-D objects. Spatial reasoning skills are acquired in early childhood where often haptic and visual experiences are combined to build 3D mental models of the real world. Since high-D objects are largely mathematical and do not occur in a tangible form, the associated cognitive reasoning chains are not developed in these critical early years. This lack of reasoning faculties represents a barrier for most people when dealing with high-D data later in life and so deprives them of the chance to find more insight in these data.

We describe a framework and interface that eases this barrier by design, called the *Subspace Voyager*. It serializes the exploration of high-D space into a continuous travel along a string of generalized, but not necessarily dimension axis-aligned 3D subspaces, visualized as scatterplot projections of the data points. This serialization allows us to abolish the complex interactions and representations that are often typical to high-D space exploration tools. We replace them with paradigms familiar to most people, such as trackballs, maps, and word clouds. Our interface uses these to help users explore the generalized 3D subspaces, navigate the continuum of 3D subspaces, and assess the relevance of individual attributes for a given subspace.

The simplicity gained through the 3D subspace decomposition comes at a price – the extent of the transformations defined on such a restricted subspace is limited and may not reach far enough to generate a projection in which a pattern of current interest is well expressed. To enable a reach beyond these limits we have augmented the 3D navigation interface with extra capabilities that allow users to "chase" the discovered patterns by moving to adjacent 3D subspaces via simple mouse interactions. In this way, patterns can be observed that are truly multivariate and not restricted to a single 3D subspace.

In some sense, our approach is akin to that taken in an upcoming Indie video game, *Miegakure* [46] (itself inspired by the classic novel *Flatland* [2]) which enables 4D space travel by swapping one of the three current dimensions. We go significantly further than this game: (1) our spaces are much greater than 4D, and (2) we allow transitions in all dimensions simultaneously. Yet, it is encouraging that the entertainment industry sees fun in this type of space travel. It suggests that our interface might be fun and engaging as well, which will immensely benefit the analytics that is performed with it.

The 3D subspaces our system supports are general in the sense that they do not need to be constrained to three specific data axes but

• *Bing Wang and Klaus Mueller are with the Visual Analytics and Imaging Lab at the Computer Science Department, Stony Brook University, Stony Brook, NY. Email: {wang12, mueller}@cs.sunysb.edu.*

can be spanned by a basis of three arbitrary orthogonal vectors. This affords a better alignment with the high-D phenomenon under study and effectively allows its exploration in relation to all relevant data dimensions. It, however, also brings about a huge number of possible subspaces. To manage this complexity we provide a variety of objective-driven search and clustering facilities that assist users in locating subspaces with interesting structures.

When designing our interface we placed great emphasis on making the interactions direct, intuitive, and responsive [34]. Most exploration goals can be achieved by expressing them directly in the visualization, via simple mouse selections and transitions. At the same time, our framework is quite general and is readily applicable for many tasks and application areas that involve multivariate data, such as cluster sculpting [30] and analysis, information discovery, and the supervised training of classifiers, just to name a few.

In summary, the specific contributions of our work are:

- A serialization of high-D space exploration into a journey within and across a string of adjacent generalized 3D subspaces
- An interactive trackball interface for 3D subspace exploration augmented with direct controls for goal-directed transitioning to adjacent 3D subspaces – an activity we call *cluster chasing*
- An illustrative, non-obtrusive labeling scheme that allows users to appreciate the influence of different variables on the display
- Various goal-directed view optimization and view selection facilities that lower the subspace navigation overhead and expand the search for interesting high-D phenomena
- A map-like interface organized by view similarity where users can store interesting scatterplot views and construct a tour for presentation within an animated scatter plot display

Our paper is organized as follows. Section 2 reports on related research motivating our work. Section 3 focuses specifically on the TripAdvisor$^{ND}$ system – a precursor of the Subspace Voyager. Section 4 provides a system overview. Section 5 describes the trackball based subspace explorer. Section 6 presents the subspace trail map. Section 7 outlines three use scenarios. Section 8 describes our user study and its outcomes, and Section 9 offers conclusions.

## 2. Related Work

Our principal visualization modality is the scatterplot – a projection of the data into an orthogonal 2D basis. In scatterplots, clusters and their shapes are relatively easy to recognize, but points distant in high-D space may project into similar locations and this can lead to ambiguities. Helping users deal with these ambiguities is one of the major aims of our work. Another aim is to aid users in producing informative projective views into interesting subspaces of the data. In the following, we divide work related to ours into four aspects.

### Dealing with projection ambiguities

One way to resolve projection ambiguities is to decompose the space into a matrix of axis-aligned bivariate scatterplots, called SPLOM [17]. While SPLOMs can help with disambiguation, users might find it difficult to integrate information from such a mosaic of plots, especially when the number of dimensions is even moderately large.

Another approach is to use layout optimization schemes, such as Multidimensional Scaling (MDS) [24], Linear Discriminant Analysis (LDA) [28], and Stochastic Neighbor Embedding (t-SNE) [42]. MDS, for example, seeks to generate a layout where the pairwise distances of points in 2D are relatively similar to those in high-D space. But even with layout optimization, trying to warp high-D space onto a 2D plane is inherently ill-posed since it cannot fully capture multivariate data variations. Distortions are the consequence, making it difficult to correctly recognize the true shape and appearance of clusters, and also hampering the assessment of point-wise distances, both far and near. Hence, while ambiguities might be resolved, the risk of distortions has taken their place.

A third alternative is to enable users to change the projection basis in a continuous fashion, effectively using motion parallax to resolve depth and relative distance. Several systems have followed this paradigm. One of these is ScatterDice [11]. It restricts the transitions to motions between two bivariate projections at a time, giving rise to a dynamic 3D point-cloud projection display. More general is the GGobi system [39], itself derived from the seminal concept of the 'Grand Tour' [5], as well as the TripAdvisor$^{ND}$ framework devised by one of the co-authors [31]. Both allow users to transition between arbitrary multivariate projections. Our current framework also follows this general paradigm but offers interactive exploration capabilities that greatly exceed those provided by these earlier systems. For example, while GGobi also uses a trackball, it does not offer the advanced subspace exploration facilities our trackball interface provides.

### Defining a multivariate projection basis

Our layout is a generalized projection display where the 2D location $\boldsymbol{p}$ of a projected $N$-D data point $\boldsymbol{x}$ with coordinates $x_i$, $0 \leq i \leq N\text{-}1$, is given by $\boldsymbol{p} = \sum_{i=0}^{N-1} \boldsymbol{v}_i\, x_i$. Here, the $\boldsymbol{v}_i$ are a set of 2D basis vectors with common origin $\boldsymbol{O}$. We can use this formulation to compare our display with several others that are in common use. In Star Coordinates [21] all basis vectors have unit length and by ways of changing their orientations, users can interactively increase the spread of the projected data points. RadViz [20] is similar but includes a normalization by $\sum_{i=0}^{N-1} x_i$. Conversely, in biplots [14] the vector basis is a projection of the axis vectors into the 2D frame spanned by the two major principal component (PC) axes. As a result, the vector $\boldsymbol{v}$ are typically not (all) unit length and their orientation is clearly defined. Projecting the data points into the PC-basis naturally maximizes their spread in the 2D display which removes the need for interaction. However, the projection ambiguity problems still remain.

Our display is similar to biplots but distinct in two ways. First, we allow users to change the biplot projection basis interactively which helps overcome the ambiguity problems via motion parallax. The transitions can affect many dimensions at once, and not just one at a time like in Star Coordinates and RadViz. Second, we plot the dimension labels at the display periphery. We use the sizes and opacities of the dimension labels to indicate the influences of the attributes on the projection. Conversely, biplots project the data axes as arrow-headed lines directly into the display leading to clutter.

### Selecting informative views

The problem of projective view overload is not unique to SPLOMs. In many cases, it can be helpful to include proper quality criteria by which the most informative views can be selected. Research in this area has mainly addressed the selection of axis-aligned views in the presence of clustered or classified data. Sips et al. [37] define a class consistency measure which favors views based on the distance to the class center of gravity or on the entropies of the spatial distributions. Tatu et al. [40] assess quality by measures on density, histogram, and class separation. The rank-by-feature system [36] allows users to specify certain statistical criteria, such as correlation, scatterplot uniformity, etc. Schäfer et al. [35] describe a quality metric that focuses on structural preservation and visual clutter avoidance. GGobi uses projection pursuit [9][13] to generate interesting multivariate projections. We use a popular evolutionary algorithm – ant colony optimization (ACO) [10] – in conjunction with view quality metrics such as stress, class density, class separation, holes, and central mass.

Finally, a problem with having many projections is also how to manage and organize them. Several map-based diagrams have been proposed [31][45]. We provide a novel map that is dedicated to the management of generalized subspaces. In addition, our map also allows users to construct animated tours for presentation purposes.

### Managing interesting subspaces

Subspace clustering has been an active research area in the data mining community [23] but the focus was mostly on automated algorithms. In the field of visualization, one may distinguish the contributions by how much they rely on automated subspace analysis methods. On one end are the works of Yuan et al. [44] and Kim et al. [22] where users are in full control. The former proposes a visual subspace exploration approach that focuses mainly on interactive dimension set selection and refinement. The latter suggests a system where users can drop data points into two different groups and the projection basis vectors are updated automatically. Lehmann et al. [26] find minimal sets of projections, allowing users to draw a path to traverse between them. In our system, users can also modify the projection basis to favor certain dimensions, namely by emphasizing the influence of these dimensions directly in the interface.

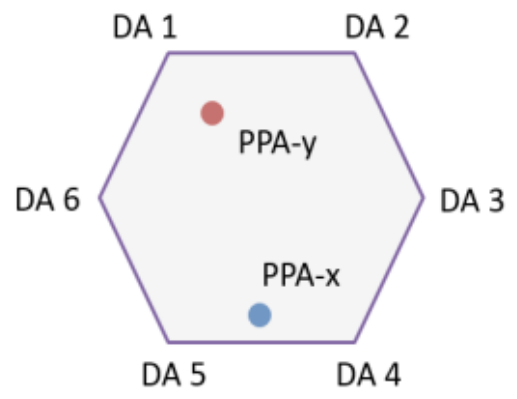

Fig. 1. Pad-based navigation interface of TripAdvisorND. In the setting shown, the PPA-x vector is dominantly a combination of dimension axis DA 4 and DA 5, while PPA-y is a combination of DA 6, DA 1, and DA 2.

Other approaches first perform an automated subspace clustering step and then visualize the results as small multiples of scatterplot projections [4], as MDS layouts [41], or use animated transitioning between them [27] akin to our map. We also first perform clustering but then use the results to provide guidance in the subsequent visual exploration of the actual subspaces, focusing on cluster appearance and relations. This can be helpful in the visual reasoning process.

Related in some respect is also the LineUp system by Gratzl et al. [16]. LineUp requires users to manually set a weight for each attribute to determine its influence on the rankings of the data items. However, setting weights explicitly might not be intuitive to mainstream users with limited quantitative reasoning abilities. They may simply not know their preferences at this level of detail but rather discover them implicitly during data exploration. Our system supports this type of exploratory discovery process.

## 3. Recap: The TripAdvisor^ND Framework

The approach we have taken is largely motivated by our earlier TripAdvisor^ND framework [31] and the shortcomings we have observed in its use. One major improvement is the new trackball interface, which is much more direct than the spatially disjoint navigation pad of TripAdvisor^ND (see Fig. 1). This navigation pad consists of a polygon with *S* vertices, where *S* is the cardinality of the subspaces. Each vertex corresponds to a native dimension – hence the subspaces are axis-aligned (and not generalized). It should also be noted that for S>3 different orderings of the vertices are required to allow users to access the full projection coverage of the subspace.

The interior of the polygon shows two disk-shaped pointers. They represent the two (N-D) basis vectors into which the N-D point cloud is projected for display using the vector dot product. In [31] these two vectors are called *Projection Plane Axis (PPA) vectors* – the x-axis is PPA-x and the y-axis is PPA-y. The vectors are computed from their positions in the pad polygon via generalized barycentric coordinate interpolation [29].

In the pad-based interface, users can control the influence a dimension has on the display by moving either the PPA-x or PPA-y pointer toward that dimension. This essentially spreads out the projected point cloud along that dimension and so reveals the dimension's ability to separate the data points into different populations/clusters. Then, by moving the other pointer toward another dimension, bivariate relationships can be visualized. Finally, when moving either or both pointers midway between a set of dimensions users can appreciate the combined effects stemming from the multivariate relationships of these dimensions.

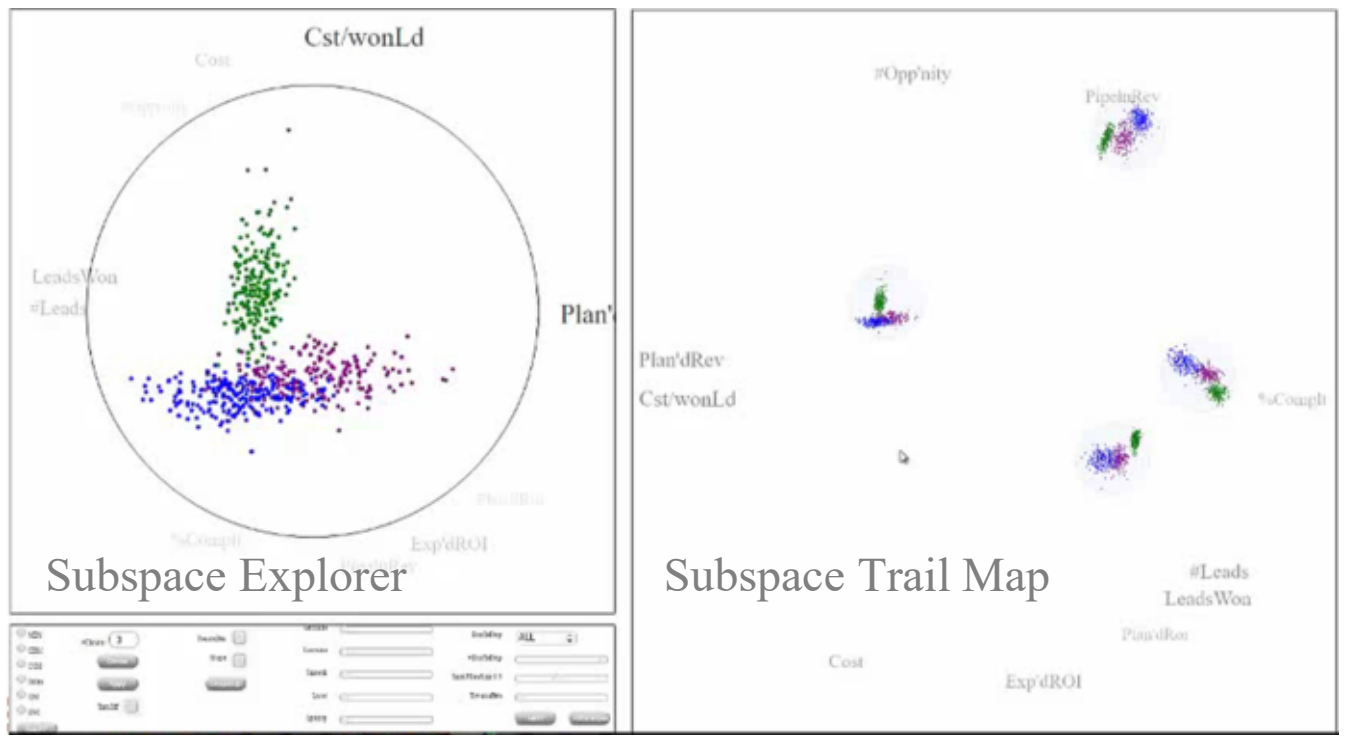

Fig. 2. Subspace Voyager interface. It has three main components: the Subspace Explorer (SE), the Subspace Trail Map (STM) and the control panel. The SE is coupled with the trackball interface. It not only displays the data as a scatterplot, but it also allows users to visualize the current directions of the projected dimension axis vectors as labels placed outside its circular boundary. The labels are properly sized in terms of the corresponding attribute's influence on the display. The SE offers various interactions for users to examine the data. The STM holds a set of views (and their parameters) that users may have found interesting during the exploration, embedding them into a word cloud of attributes. Finally, the control panel allows users to set the various parameters and modes in the system.

**Shortcomings of TripAdvisorND motivating our work**

While the pad interface allows unprecedented control in the dynamic manipulation of the view onto the N-D point cloud, the need to separately manipulate two pointers in sequence suffers from a certain lack of ergonomics. A further shortcoming is that users are required to keep track of two interfaces at the same time: (1) the visualization window that shows the moving point cloud along with a projected coordinate system, and (2) the pad that controls the orientation of the projection plane. In practice, a user may observe one or more dimensions that should be emphasized in the display as they might offer the potential to break up a cluster into two or more components. To do this, the user would need to looks at the pad to identify which pointer should be moved and in what direction, and then observe the effect in the display. In the present work, we aimed for an interface that makes this operation more straightforward by embedding the navigation controls directly into the display. Enhancing the well-known trackball interface with N-D navigation capabilities seemed to be good choice toward this goal. We also added view optimization and other navigation aids to support the manual exploration, allowing users to arrive at meaningful projections faster.

## 4. System Overview

Fig. 2 shows the Subspace Voyager interface. It has three main components: the *Subspace Explorer (SE),* the *Subspace Trail Map (STM),* and the control panel. The latter allows users to set the various parameters and modes in the system.

The exploration pipeline of the Subspace Voyager is illustrated in Fig. 3. After loading the data, our system performs either Random Projection or Subspace Clustering and Principal Component Analysis (PCA) to identify an initial promising 3D subspace. More 3D subspaces can be generated via the control panel at any time.

The data is then projected into this generated subspace and is displayed in the SE-embedded trackball. There are different interaction modes users can perform on the trackball. The first mode is to rotate the trackball while pressing down the left mouse button. This enables an exploration of the *current* 3D subspace. The second mode allows users to transition to *adjacent* subspaces where certain attributes of interest have a higher emphasis than in the current 3D subspace. It yields data projections that better capture the cluster distributions in these attributes. In this *cluster chasing*, users move the mouse – now with the right mouse button depressed – toward the respective attribute labels displayed on the trackball periphery. This increases the weight of these dimensions in the PPA vectors.

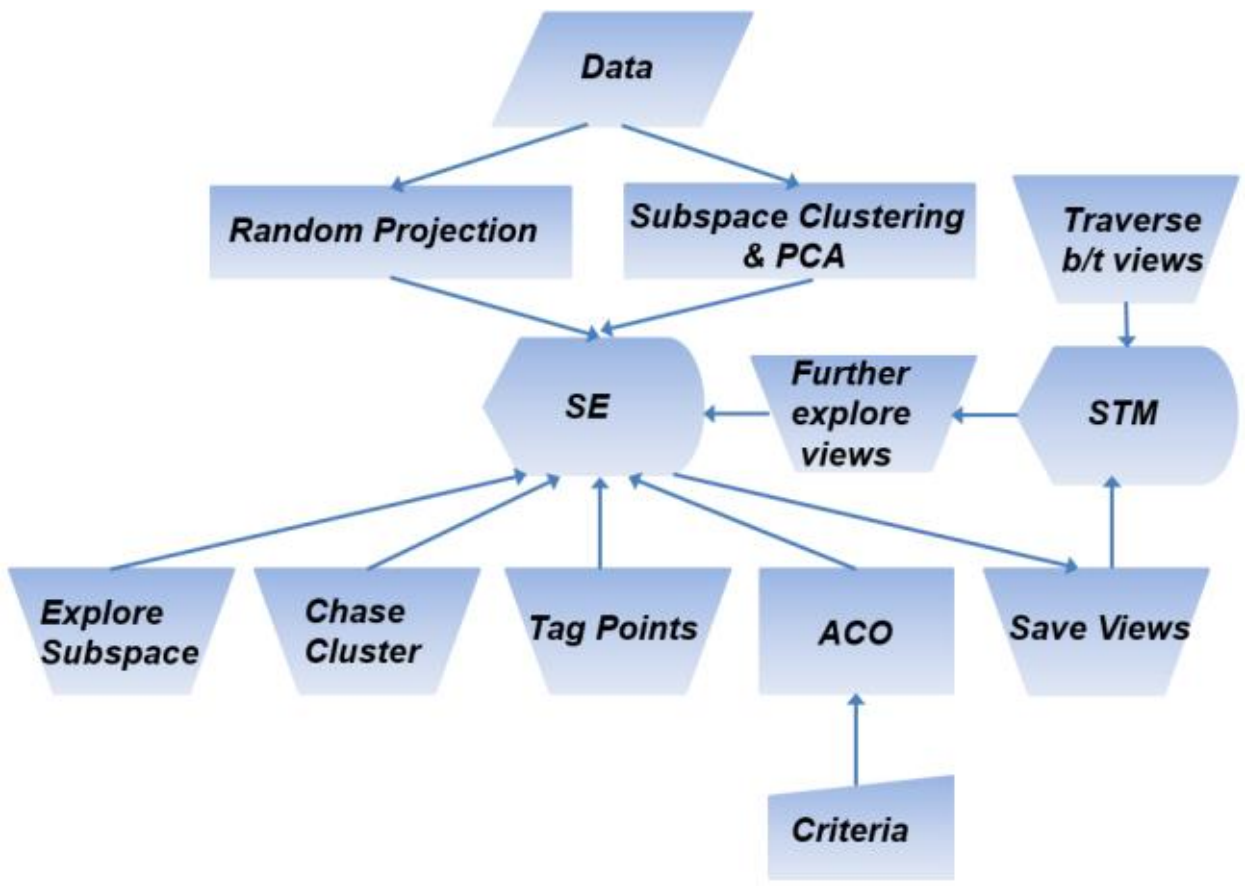

Fig. 3. *Subspace Voyager* workflow. See Section 4 for a narration.

As mentioned, in our system there is no need for manually optimizing views which can be tedious. Our system provides Ant Colony Optimization (ACO) [10] to generate the best trackball configuration automatically according to a set of user-selected view quality criteria. Users can also tag points by brushing them into different colors. This is helpful for cluster analysis or for editing out unwanted structures. Finally, at any time users can save the current trackball view to the *STM* to keep track of interesting findings. Any of these STM views can then be dragged back into the trackball for further exploration. Multiple small views can also be linked and traversed in order, providing a smooth transition between views.

### 4.1 Generating a Set of Subspaces

Choosing meaningful subspaces for exploration is a key challenge in multivariate data analysis and much work has been dedicated toward this goal (see Section 2). We have implemented two such strategies: (1) random view generation and (2) subspace clustering. Users can generate new subspaces at any time via the control panel.

For the former (1), we use the technique proposed by Anand et al. [1] and then further optimize the subspace using ACO powered projection pursuit (see Section 5.4). For the latter (2) we assume – similar to Liu et al. [27] and our own work [43] – that each cluster forms a subspace on its own. We characterize each such subspace by the three principal components obtained with PCA. Finally, for both of these methods, we use ACO view optimization to generate a high quality (given the chosen metric) scatterplot projection in the trackball display.

We should also note that in a view that has the PC vectors as its basis if two (or more) dimension vectors are very close, it means they are to some extent correlated. This is especially true when these dimensions have large weightings in one significant PC (i.e. these dimensions are strongly correlated [47]). We will make use of this relationship in the use case described in Section 7.1.

### 4.2 The Subspace Explorer (SE)

The SE is coupled with the trackball interface. It not only displays the data as a scatterplot, but it also allows users to visualize the current directions of the projected dimension axis vectors as labels placed outside its circular boundary. The size and opacity of a label indicate to what extent its associated attribute is expressed in the projection. A larger and bolder font means that the scatterplot exhibits more of the attribute's variability. The label placement, on the other hand, reveals the radial direction along which the variability is mostly exposed.

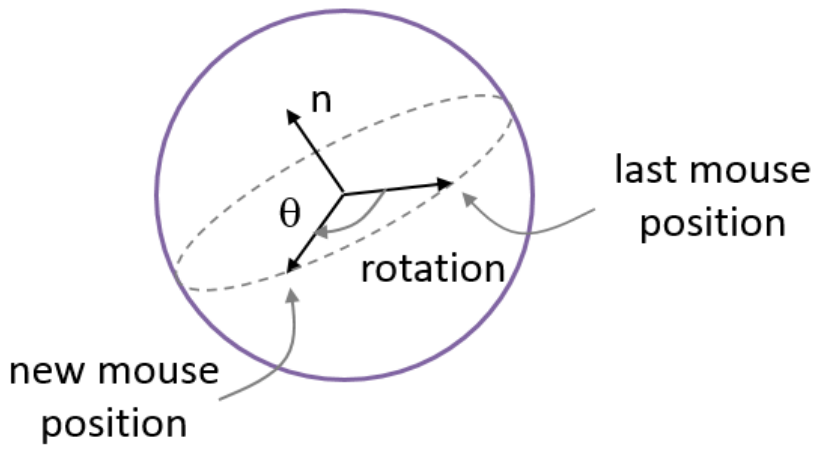

Fig. 4. 3D trackball. Given the current and previous mouse clicks, both the axis of rotation and the rotation angle can be computed.

The simplest form of trackball interaction generates scatterplot projections confined to the current generalized 3D subspace projected into the SE. This projected 3D subspace can be modified by:

- Mouse-initiated trackball interaction: users can transition to adjacent 3D subspaces by augmented trackball interaction
- Randomized projections: this discovers new 3D subspaces ready for trackball-based exploration
- 3D Subspace interpolation: moving a slider in the control panel generates a continuous set of 3D subspaces, intermediate to two subspaces in the STM, which can be explored via the trackball
- View optimization: the 3D subspace (as well as the current projection view within the current 3D subspace) can be optimized via projection pursuit driven by a user-defined set of criteria

The control panel provides several options for trackball use. The checkbox 'TurnOff' specifies if all data points are to be shown or only those that are well described in the current subspace, i.e., belong to that subspace. The color bar on the bottom right is the brushing tool. It allows users to tag individual points or groups of points in a dedicated color to cluster them or mark them as inactive in gray.

### 4.3 The Subspace Trail Map (STM)

The STM holds a set of views (and their parameters) that users may have found interesting during the trackball exploration. The view images are embedded into a word cloud of attributes. Their placement with respect to each word indicates the influence of the corresponding attribute to the view. We treat each view as a point and use PCA on all of them to spread them out. The circular shape of the images mimics the shape of the trackballs. A smaller diameter reduces overlap of similar views in the STM while a larger diameter provides magnification. Users can drag any view back into the trackball for further exploration, or they can connect interesting views by lines to produce animated transitions for presentations.

## 5. The Subspace Explorer and Trackball Interface

Users can tilt the trackball and watch the resulting scatterplot react to the motion. Fig. 4 sketches how a trackball works. Imagine a virtual sphere that encapsulates the current generalized 3D subspace. When clicked, the screen coordinate of the mouse is mapped to this sphere. Given the current and previous mouse clicks, we can compute the axis of rotation $n$ and the rotation angle θ. From those two quantities, a 3×3 rotation matrix is derived, as described in [3].

### 5.1 Creating the Trackball Space Projection Matrix

The trackball system only works in 3D but our data points are N-D and so we need to project the ND points into 3D before rotating. We achieve this by post-multiplying the trackball rotation matrix T with the 3×N projection matrix P. We have two options for the first two of the vectors in P: (1) the orthogonal PPA x-axis and y-axis pair we obtained from the randomized projection procedure, or (2) the two most significant PCs we obtained when performing PCA for the selected cluster. In both cases we require a third orthogonal axis, call it the PPA z-axis. Since this is N-D space we have a number of choices. We can either (1) randomly generate an N-D vector, or (2) if the PPA x-and PPA y-axes are generated via PCA, use the third most significant axis for the PPA z-axis.

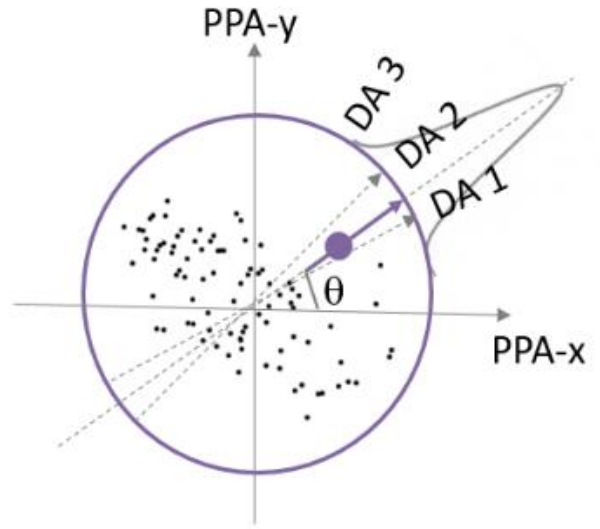

Fig. 5. Updating the PPA x-axis and PPA y-axis vectors by moving the mouse towards one or more dimensions. The influence of each dimension is weighted by a Gaussian function.

Note that the resulting vector is not necessarily orthogonal to the PPA x-axis and the PPA y-axis. To make it orthogonal we use the Gram-Schmidt orthonormalization process [7] to find orthogonal basis vectors. The Gram-Schmidt process takes *N* linearly independent vectors and produces *N* orthonormal vectors spanning the same N-D space. In practice, we keep the PPA x-axis and PPA y-axis which are already orthonormal and run Gram-Schmidt to orthonormalize the PPA z-axis from the initially chosen vector. Once P is configured in this way, T is reset to the identity matrix, ready to be manipulated in the 3D trackball interaction.

### 5.2 Processing the Points within the Trackball Space

With P in place, the following sequence of operations is executed for every trackball move: (1) compute the 3×N compound projection matrix M=S·T·P, where S is an optional scaling matrix that allows zooming into the display, and (2) multiply each N-D point vector $V^{ND}$ by M to obtain the 3D points $V^{3D}=M \cdot V^{ND}$. But ultimately we are only interested in the projection of the points into the coordinate system spanned by the PPA-x and PPA-y vectors manipulated with the trackball. This yields a set of 2D points, $V^{2D}$, which are the first two components of $V^{3D}$ since the projection is orthogonal.

We have not observed a significant delay in the direct projection of N-D points in the operation of the trackball. But first pre-computing a 3D point cloud right after construction of the 3D coordinate system and rotating them directly for the lifetime of P can reduce the number of computations to roughly N/3 of the original computations. We have not chosen this intermediate step because it requires extra storage which can be significant for large point clouds.

### 5.3 Mouse Interactions within the Trackball Interface

We provide three modes of mouse interactions within the trackball. All are controlled with different mouse buttons depressed. The first is the basic mouse interaction when the trackball is rotated within the current 3D subspace. It is performed when the left mouse button is depressed (see Section 5.2). The other two operations are described in more detail in the following.

#### 5.3.1 Chase clusters in adjacent 3D subspaces

When using the basic 3D subspace exploration mode (Section 5.2) we frequently observed that interesting patterns were starting to evolve but their full exposure was out of reach since it occurred in a different, albeit nearby, subspace (i.e. a subspace that could not be reached simply by 3D rotation). In these situations, we often felt the need to "break out" of the current 3D subspace in the direction of the trackball movement such that these patterns could be reached. To solve this shortcoming we added the capability to smoothly transition from one subspace to an adjacent one. It allows users to interactively change the influence of the data dimensions whose projections align with the current trackball motion, progressively increasing their bias in the projection matrix P. This gives the exploring user access to the adjacent 3D subspace where the patterns of interest are better expressed. It lets him/her explore the data with a higher emphasis on one or more attributes of interest.

To engage into this mode of exploration users would release the left mouse button and instead press the right button while moving the mouse in the direction of the desired dimension's projection, as indicated by the corresponding attribute's label on the trackball's periphery. The further the mouse is moved the more the projection plane is tilted into the dimension's axis vector. Conversely, moving backward along that direction, towards the center of the trackball, decreases the influence of this dimension.

As Fig. 5 illustrates, ideally we would accomplish this task by adding (or subtracting) increments $\Delta x=k_a \cdot \Delta d \cdot \sin(\theta)$ and $\Delta y=k_a \cdot \Delta d \cdot \cos(\theta)$ to the PPA-x and PPA-y vectors, respectively, where θ is the angle between the mouse movement vector and the trackball x-axis (the PPA-x vector). Here Δd is the distance the mouse moved in the direction of the projected dimension axis vector (positive when moving towards the periphery, negative otherwise), and $k_a$ is a user-adjustable speed constant (we use the dot products instead of the trigonometric functions). Subsequently, Gram-Schmidt is used to re-orthonormalize P (see Section 5.1), using the original PPA z-axis vector. One problem here is that, after Gram-Schmidt, the direction of this data dimension would change and thus there might be other dimensions taking the selected one's direction. We overcome this by fixing the selected dimension until the user releases the mouse.

This basic approach generalizes to more than one dimension. Fig. 5 illustrates the practical case in which there are two or more projected dimension axis vectors in close range of the exploration direction. This might be an indication of multivariate relationships. To properly scale the axes vector influences geometrically, we apply a Gaussian weighting in terms of their direction misalignment. This is done via the following equation: $w_d=\exp(-k_d \cdot \mathrm{dot}(v_m, v_d))$ where $w_d$ is the weight applied to this axis vector, $v_m$ and $v_d$ are the direction vectors of the mouse and the axis vector, respectively, and $k_d$ determines the reach of the Gaussian. The remaining steps are similar to the single-vector case described in the previous paragraphs.

Our system also supports the case in which a user would first select an attribute via a mouse click on the trackball boundary but then move the mouse in a direction not necessarily aligned with the attribute's dimension vector. This will gradually align the dimension vector with the mouse motion and move the attribute label accordingly. Again, the selected dimension's weighting changes according to the direction and length of the mouse movement.

#### 5.3.2 Go "deeper" into high-dimensional space

By clicking the middle mouse button, our system generates a PPA-z vector according to the two options described in Section 5.1 and a new orthogonal vector is computed using Gram-Schmidt. Then with a trackball up (down) motion, the emphasis of the dimensions projecting on the PPA z-axis is increased (decreased). The effect of this operation will only be visible once the trackball is rotated regularly and the new 3D subspace is exposed. We call this functionality "deep" since the axis that is changed is the PPA z-axis (i.e. the axis pointing into the depth of the display).

### 5.4 Display of Attribute Labels on the SE Boundary

As mentioned, in order to better comprehend the relationships between a scatterplot projection and the data dimensions (attributes), we display the attribute names as labels along the SE trackball periphery (see Fig. 6). The extent of which a dimension contributes to the projected point cloud is indicated by label size and opacity. The larger and bolder the label's font is, the stronger the attribute's contribution to the plot. The location of each label is computed by the attribute's weighting in the PPA-x and PPA-y vectors. Let $w_x$ be the PPA-x weighting, and $w_y$ be the PPA-y weighting. Then the angle between this dimension vector and the positive x-axis is computed as $\alpha = \mathrm{atan}(w_y/w_x)$.

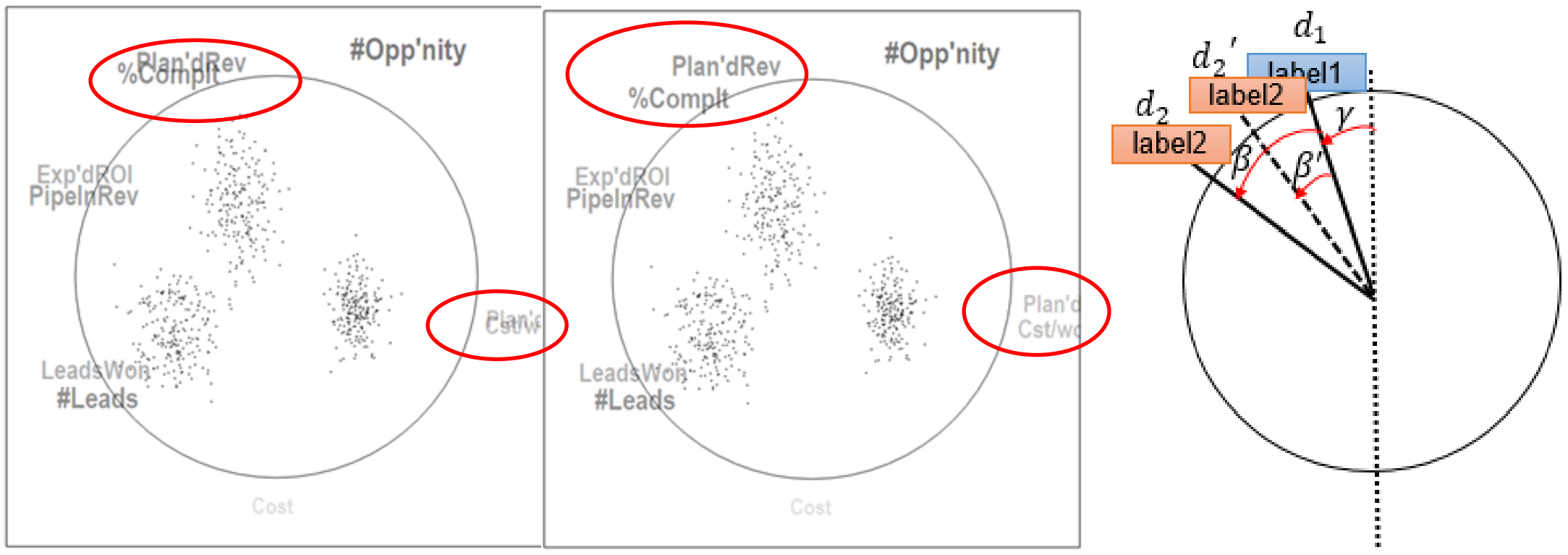

(a) (b) (c)

Fig. 6. Dimension label overlap prevention. (a) Naïve implementation causing label overlap; (b) Using our angular spacing scheme to prevent label overlap; (c) Illustration of our label overlap prevention scheme.

#### 5.4.1 Preventing Overlapping Attribute Labels

In practice, attribute labels may come to print on top of one another (Fig. 6(a)). This occurs because several dimension vectors overlap. We solved this problem by forcing labels to locate at least β degrees apart from their neighbors. Fig. 6(c) shows this for the upper left quadrant. Here, $d_1$ is the location of label1 located γ degrees away from PPA-y and $d_2{}'$ is the location of neighboring label2, spaced $\beta'$ degrees away. We see that $\beta'$ is too small causing the two labels to overlap. Therefore we introduce a small displacement which places label2 at $d_2$. Now label1 and label2 are spaced $\beta$ degrees apart and no longer overlap.

In experiments, we found that the best choice for β is dependent on the orientation of the dimension vector. The more vertical it is, the larger β should be, while for a more horizontal alignment, a smaller β will suffice. The following equation relates β to the angle γ between the vertical axis and the dimension vector (for the upper left quadrant only – the other three quadrants are related by symmetry):

$$\beta = \begin{cases} \theta_v - (\theta_v - \theta_h) * \frac{\gamma}{45°} & 0 \leq \gamma < 45° \\ \theta_h & 45° < \gamma \leq 90° \end{cases}$$

Here, $\theta_v$ and $\theta_h$ are constants we determined for the maximal font size of the labels which occur when the corresponding dimension vectors are fully projected. The angle $\theta_h = 4°$ is the displacement needed when γ is greater than 45°, while an angle of $\theta_v = 24°$ is needed when γ=0°. When γ is between 0° and 45° we determine β via linear interpolation. Fig. 6(b) shows the configuration of Fig. 6(a) with our label displacement scheme enabled.

We also found that while displacing the labels provided for better readability, it was distracting in interactive mode when users were rotating the trackball since it could lead to sudden jumps of the labels. Hence we only apply the overlap removal method when the projection is fixed (after releasing the mouse). Conversely, when a dataset has many dimensions, the label overlap can never be prevented. For this reason, we added a slider to the control panel by which users can set the maximum number of displayed attribute labels. These can be the most significant attributes or attributes manually selected by clicking on their labels with <ctrl> depressed.

### 5.5 Point Brushing, Tagging, and De-Activation

Our interface also provides the ability to label a point (or a group of points) with a color chosen from a palette. This is useful when monitoring a certain point's (or point group's) behavior when the trackball rotates. It greatly helps in distinguishing different clusters or seeing sub-clusters emerge during motion.

Conversely, by painting a selected group of points in gray they will become invisible and will be excluded from all further analysis. This helps, for example, in recognizing other structures that were hidden or ambiguous before this removal.

## 6. The Subspace Trail Map and View Generation

The subspace trail map (STM) is a spatial layout of thumbnail representations of views. It serves three purposes. First, it enables users to keep track of the subspaces explored so far. These subspaces can be revisited for further exploration. Second, it serves as a presentation platform for the system to suggest new subspaces not yet explored. Third, it permits users to define routes along which they can transition between two or more of these subspaces, essentially using them as keyframes. In the STM, users can double click any view thumbnail and add it back into the SE. For clustered data, all subspaces can be inserted into the STM at once by clicking the 'AllSubspace' button in the control panel.

### 6.1 Populating the Subspace Trail Map (STM)

Each view thumbnail in the STM holds the view's 2D scatterplot embedded into a circle to mimic its appearance in the SE. PCA analysis is used to ensure a well-spread layout of the view thumbnails with a minimum of overlap. If overlap occurs the 'SmallViewSize' slider can be employed to lower the circle sizes uniformly (see Fig. 11(i)). Alternatively, clicking on a partially hidden view will bring it to the foreground.

To illustrate how the STM layout works, suppose there are $p$ subspace views stored in the STM and the dimensionality of the data set is $N$. The three orthogonal PPA vectors (the PPA x, y, and z-axes) spanning a subspace $j$ can then be formally expressed as:

$$PPA_{ij} = \sum_{k=0}^{N-1} w_{ijk} d_k$$

where $i$ is either $x$, $y$, or z, $0 \leq \mathrm{j} \leq p-1$, $w_{ijk}$ is the weighting of the $k^{th}$ data dimension on $PPA_{ij}$ and $d_k$ is the $k^{th}$ dimension axis vector. We then use the $L_2$ norm to define the overall weighting of the $k^{th}$ data dimension for the $j^{th}$ subspace:

$$W_{jk} = \sqrt{w_{xjk}^2 + w_{yjk}^2 + w_{zjk}^2}$$

These weights define an $N$-D vector for each subspace:

$$S_j = [W_{0j}, W_{1j}, W_{2j} \dots W_{p-1,j}]$$

This allows us to treat each subspace as an $N$-D point. We perform PCA on this space of points. We keep the first two PCs and project all points (subspaces) into this basis. Since PCA automatically seeks to find the directions that maximize the variance of the data points, the view thumbnails will be organized in a way that reduces overlaps.

Finally, the view thumbnails are embedded in a word cloud of dimension labels (see Fig. 2). These labels are likewise placed based on this PC-basis, using the projection strength of their dimension vectors to define their sizes and opacities. To prevent clutter we only keep the labels of the ten most significant dimensions.

## 6.2 Subspace and View Optimization

We perform view optimization for several tasks. One is to produce an optimized 3D subspace from a higher dimensional subspace generated via subspace clustering. Users may also use it on the fly when interacting with the SE: (1) during exploration of a 3D subspace, and (2) for chasing clusters into neighboring subspaces. In the latter case, the view optimization can be set to perform the search within a narrow range of dimension increments, or across an expanded range. Both of these applications aid users in the trackball-based exploration. They help accelerate the tedious manual exploration needed to find a view that fits a certain view quality criterion, such as a cluster or a class separation.

### 6.2.1 View optimization via ant-colony optimization

A popular view optimization method in the context of high-D data visualization is projection pursuit. Starting from any projection, projection pursuit returns the PPA x-axis and PPA y-axis that optimizes a targeted projection pursuit index (PPI). A number of methodologies have been proposed for this task, such as hill climbing [8], random search [32], or simulated annealing [9]. We have strived for a sophisticated yet comparably easy-to-implement algorithm – Ant Colony Optimization (ACO) [10]. To the best of our knowledge, ACO has not been used for projection pursuit so far.

**General description of ant colony optimization (ACO)**

ACO simulates the behavior of ants in nature. When looking for food, ants initially travel randomly until they find food. On their way back they leave a pheromone trace along the route. Instinct prescribes that other ants most likely follow this pheromone trace instead of wandering randomly. But pheromone also evaporates gradually, and so over time, shorter (lower cost) paths will be traveled more frequently and become more attractive, leading to a convergence on the optimal path. Based on this intuition, the simplest ACO algorithm consists of the following three steps executed iteratively: (1) construct solutions, (2) evaluate solutions, and (3) update pheromone, increasing it on low-cost paths and evaporating it on others. It has been shown that the solution so generated is typically quite close to the optimal solution.

The ACO algorithm requires a discrete search space. Projection pursuit, however, is typically performed in the continuous domain. General solutions that address this problem have been proposed [6][38] – we opted for a grid-based approach. In addition, ACO also requires an objective function to judge the quality of the solutions. In our case, this can be any view quality metric, no matter how complex. This freedom of choice is enabled because ACO does not require a mathematical derivation of a gradient measure which would be needed for an analytical optimization scheme.

**Specific application of ACO for subspace and view optimization**

In our case, the search space is the set of all possible PPA x-axes and PPA y-axes and the objective function is a chosen view quality metric – low stress [24], high class-consistency [37], or others. To explain how ACO works for this application, suppose (with no loss of generality) the simple case of a 2D data set with two data axes, $d_1$ and $d_2$, where the PPA x-axis and y-axis can be represented as $PPA_x = \alpha_1 d_1 + \beta_1 d_2$, and $PPA_y = \alpha_2 d_1 + \beta_2 d_2$. There are four unknowns – $\alpha_1, \beta_1, \alpha_2$ and $\beta_2$ (for an $N$-D dataset there would be $2N$ unknowns). As an illustration, these unknown parameters are represented as the four vertical gridded bars in Fig. 7.

Our ACO algorithm differs from the traditional one in the selection of the initial pheromone distribution. While the traditional ACO typically begins with an unbiased distribution, ours cannot do this since we begin from an initial PPA x-axis and y-axis configuration, e.g., a randomized view or the PC-basis of a cluster. To account for this, we increase the pheromone of this view's parameter levels, giving rise to the red path in Fig. 7, which sets its levels to the discretized $\alpha_1, \beta_1, \alpha_2$ and $\beta_2$ values of this initial view.

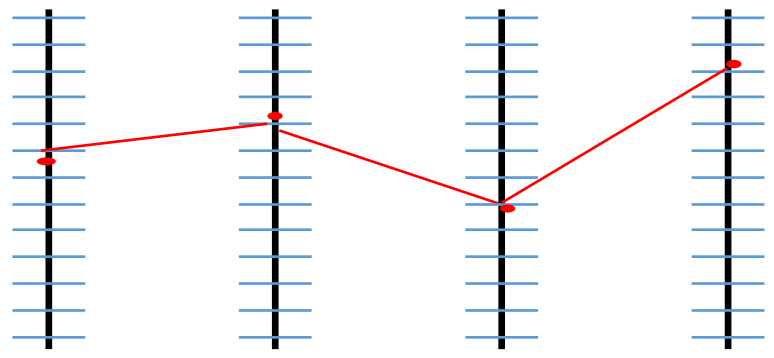

Fig. 7. Illustration of the ACO algorithm in the discrete domain. Each vertical bar grid point stands for a level of the parameter represented by the bar. The red, piecewise linear polyline is a possible solution with the levels indicated by the bar intersections.

Next, a generation of ants is set free, moving across the parameter space (from left to right in Fig, 7) selecting levels via pheromone-weighted randomization. While the levels of the initial view are more likely, the randomization ensures a more diverse set of choices. After the whole set of parameters has been traversed, the generated views are evaluated by the chosen view quality metric. The pheromone of each parameter level is then updated according to the quality of the views it was part of. The algorithm stops after a fixed number of iterations and for each view parameter, $\alpha_1, \beta_1, \alpha_2$ and $\beta_2$, the level with the highest amount of pheromone is chosen.

Fig. 7 resembles a parallel coordinate display. We observed that after the single initial polyline, ACO tends to generate many polylines which eventually narrow down to a single slim cluster – the optimized view.

ACO can also be constrained to produce views in a preferred interval. For example, one can constrain the search range on each parameter to be close to the initial path. This can be done by fixing the two ends of the vertical bars to be close to the initial values. Likewise, one can also loosen this condition and do a global search. In this case, the resulting view would be a global optimum according to different criteria. Finally, we should also take into account that the ACO needs to return PPA vectors, which are required to be of unit length and orthogonal. We therefore always normalize the returned PPA x-axis and then use Gram-Schmidt orthonormalization to find the corresponding PPA y-axis.

### 6.2.2 Other optimization capabilities

Our system also allows users to select several dimensions and produce a view in which those dimensions are equally expressed. This produces plots similar to Star Coordinates or RadViz and can be useful in cases where one wishes to see the influence of a subset of attributes on the data. It is achieved by clicking on the respective labels along the trackball while depressing the ctrl- and space keys. Then, when releasing the mouse, the weightings for the selected dimensions are set to the maximum. A Gram-Schmidt step follows to orthogonalizes the transformation matrix. Fig. 8 shows an example.

### 6.2.3 Illustrative use case

Fig. 9 shows results that can be obtained with our ACO-based subspace and view optimization framework using the sales campaign dataset described in Section 7.1. We first apply simple k-means clustering using the Structure-Based Distance Metric of Lee et al. [25] and obtain three subspace clusters. A subsequent PCA analysis for each cluster establishes three separate 3D subspaces. Clicking the 'AllSubspace' button adds all three subspaces to the STM (see Fig. 9(a)). We color the three subspace clusters blue, magenta, and green, and color the circumference of each thumbnail view by the subspace it represents. We observe that for the magenta and green subspace views, the points of the focus cluster (magenta or green, respectively) still overlap with points of other (co-) clusters – especially for the magenta subspace. Next, we optimize the three subspace views using distribution consistent criteria [37], shown in Fig. 9(b-d). We observe that the blue cluster's subspace and projection are almost unchanged. This is because the three clusters are already well separated here. Since we only run optimization in a close range of the original PC projection this view might already be the best compared to its neighbors. (Better views could possibly be obtained by expanding this range.) Conversely, the subspaces of the magenta and green clusters have significantly improved. In each panel, the respective subspace clusters are now clearly separated from the others.

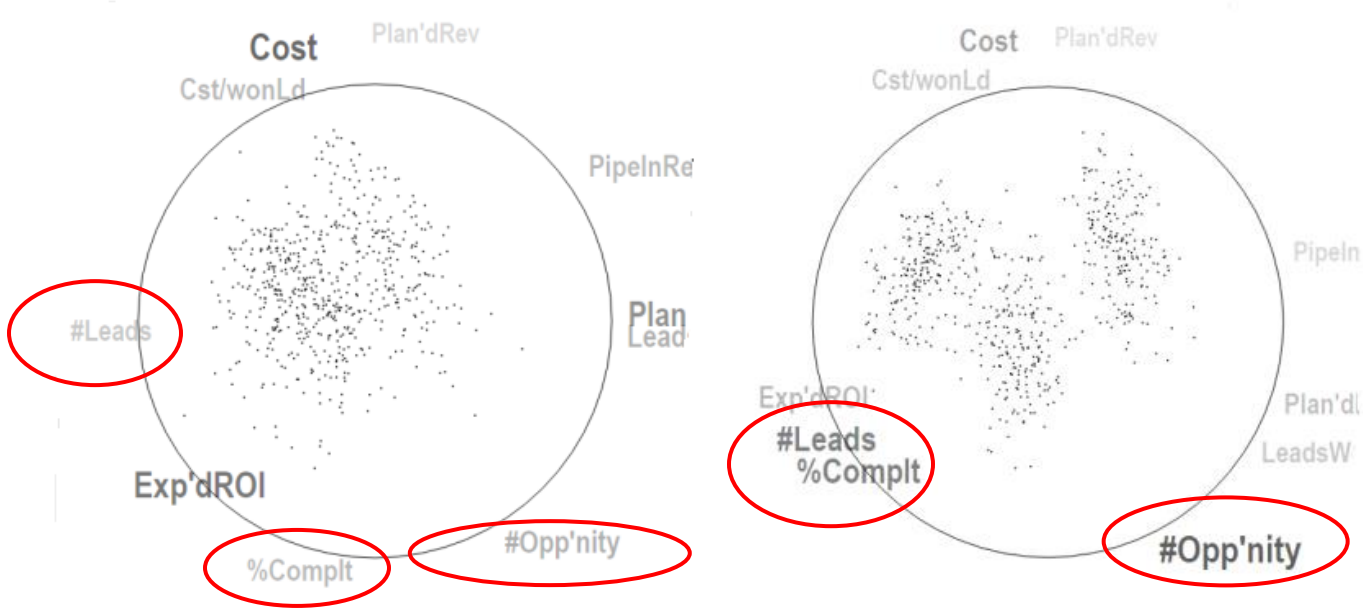

(a) (b)

Fig. 8. Equally expressing several dimensions. (a) The original projection. (b) The optimized projection where *%Complete, #Opportunity,* and *#Leads* are equally expressed.

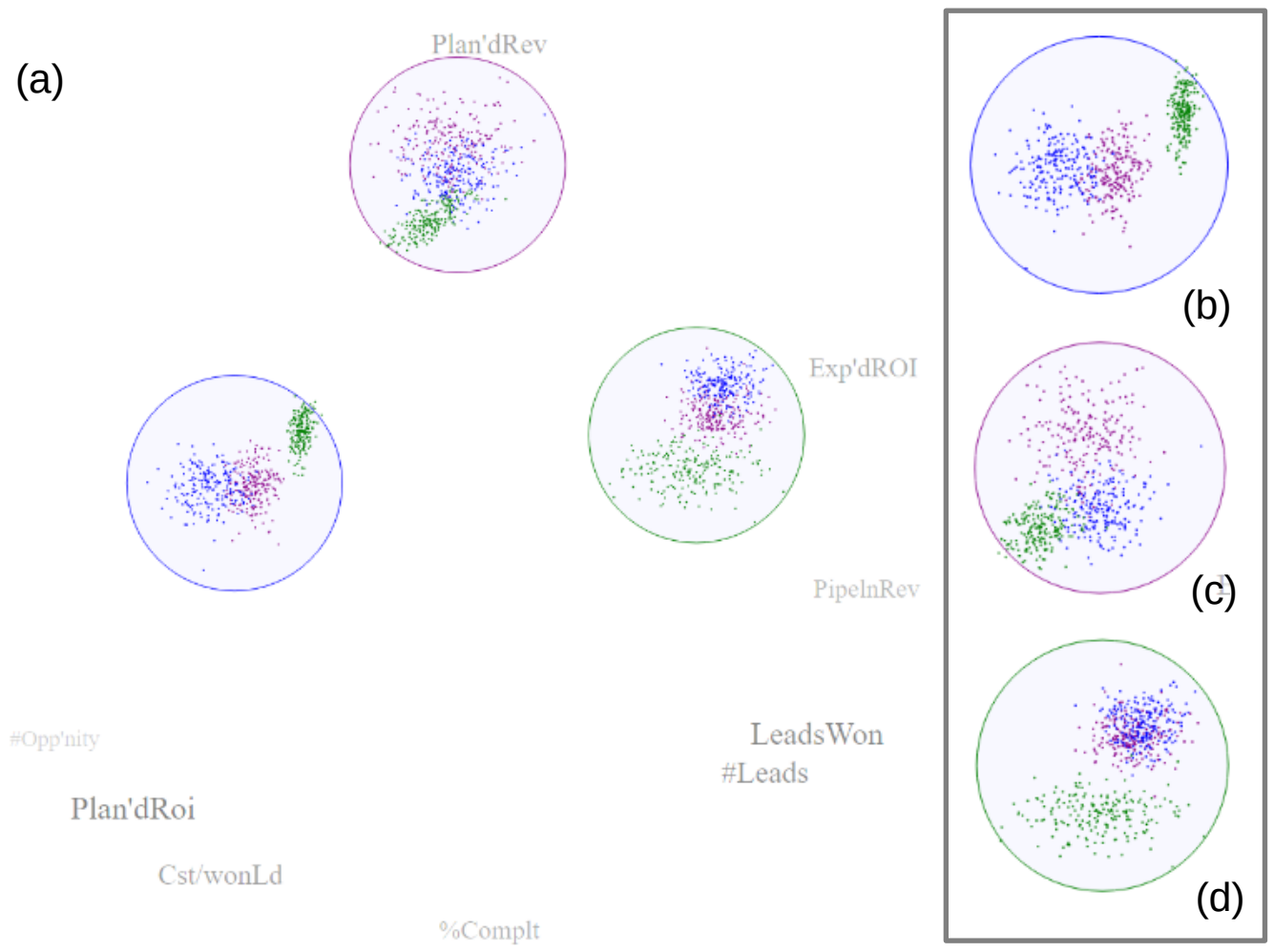

Fig. 9. Using the ACO-powered subspace and view optimizer to optimize the visual separation of three subspace clusters, colored blue, magenta, and green. (a) The STM with the thumbnail views of each subspace. The color of each thumbnail circle indicates the subspace cluster it shares its basis with. We observe that the subspace PCs alone cannot isolate the subspaces well – there is still a significant amount of cluster overlap. (b-d) Optimized subspaces for the blue, magenta and green cluster, respectively, using the distribution consistent view quality criteria. All subspace clusters are now well separated from the others in their respective subspaces.

### 6.3 Transitioning Between Subspaces

Self-initiated and controlled animation can be a helpful paradigm for humans to understand how two or more different representations of the same information relate to one another [18][33]. We have employed animation to help users understand how two subspaces relate to one another, with the added aim that this might also instill a better understanding of the high-D data space in a larger context. Users can select multiple thumbnail views in the STM and connect them with a path. Moving the 'TraverseBtw' slider then changes the PPA axis vectors from one subspace to another.

Simply linearly interpolating between bases of PPA axes, however, would lead to nonlinear intermediate projections. We, therefore, adopted the algorithm by Cook et al. [9] to transition between the two subspaces using singular value decomposition. Fig. 10 shows three snapshots of a sequence of frames from such an animation, along with the path connecting the two corresponding nodes in the STM. All keyframes and the path connecting them are shown in panel (d). Panels (a) to (c) show intermediate views along the path, and the yellow dot in panel (d) indicates the view's location in panel (b). Since these still frames can only provide a limited illustration, the reader is encouraged to view the provided video to appreciate the insightful visual effect of this animation.

Alternatively, we also include a 'presentation mode' where a narrator would click the 'Next' button to go to the next keyframe instead of using the slider. The animation provides a smooth transition between findings when presenting the results, as opposed to abruptly changing the views or simply cross dissolving them

## 7. Application Examples

In the following, we demonstrate the versatility of our framework by ways of applying it to a diverse set of use scenarios involving high-D data. We show our framework's application in (1) visual cluster analysis, (2) visual item discovery and selection, helping users to recognize and negotiate tradeoffs among items, and (3) visual cluster refinement, allowing users to partition feature-driven clusters based on the visual expression of the aggregation of these features. A fourth use case – the visual setup of a classifier in the presence of intermixing outliers – is presented in the paper's supplement.

### 7.1 Use Scenario #1: Visual Cluster Analysis

To illustrate the trackball interactions, we chose a multivariate cluster analysis task involving an interactive study of a sales force working for a large company. The dataset consists of 900 points (one per salesperson) and 10 attributes parameterizing the basic corporate sales pipeline. Briefly, a sales campaign begins with a *lead* generator who produces prospective customers that a salesperson might be able to close a deal with. If these leads receive positive responses, they become *won leads* and receive a sales pitch at a *cost per won lead*. Upon further positive response, they become *opportunities* or potential customers. *Cost* is involved in every step and high *pipeline revenue* is the ultimate goal. Three are three sales teams in our dataset.

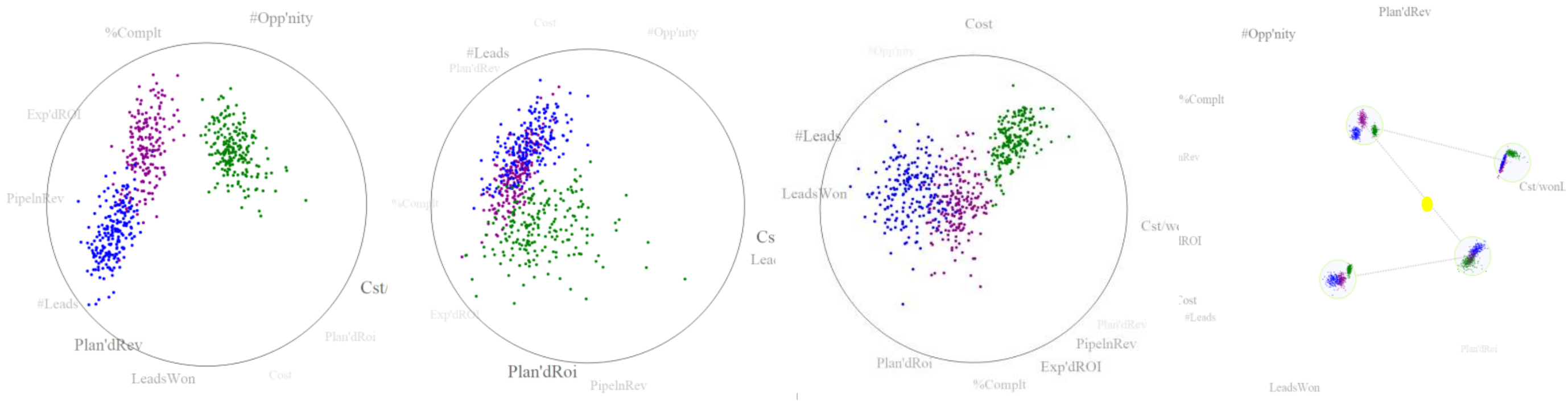

(a) (b) (c) (d)

Fig. 10. Transitioning between two subspaces marked in the STM using the animation slider. (a)(b) and (c) are three intermediate views. (d) is the animation path in the STM. The yellow dot indicates the location of the view in (b). The provided video has a complete animation.

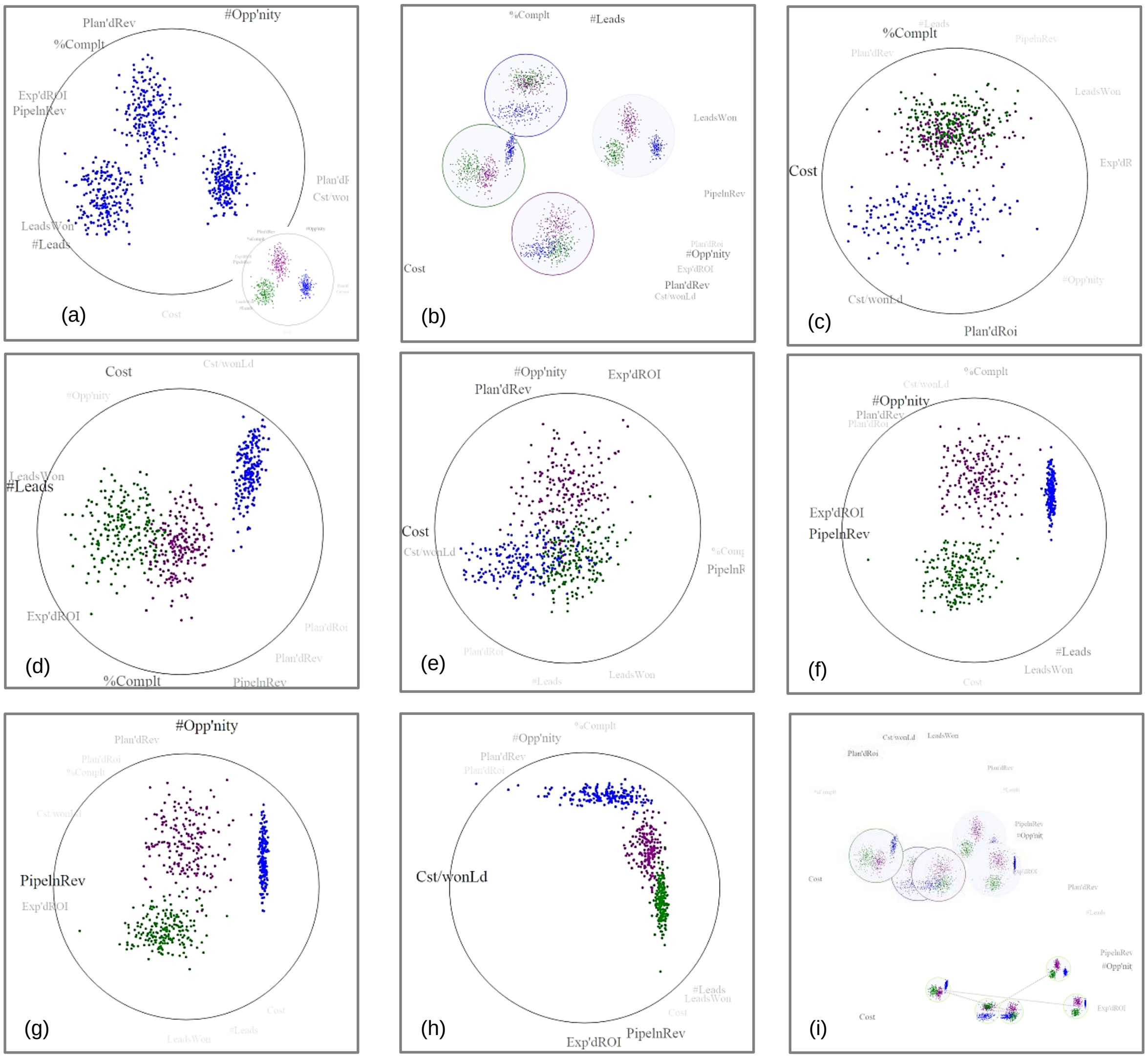

Fig. 11. Analyzing the sales force dataset. (a) The dataset projected onto the first two PCs. There are three visually separable clusters -- the three sales teams under study. (b) STM with view thumbnails of the overall space and the extracted subspaces for each of the three teams -- each optimized such that its focus cluster is maximally separated from the others. (c) Subspace of the blue team (d) green team subspace, and (e) magenta team subspace. (f) Increasing the weighting of PipelineRevenue and ExpectedROI by moving the mouse towards the respective labels (with right mouse button depressed). Both the green and magenta team generates more revenue than the blue team. (g) Increasing the weighting of #Opportunity along the PPA-y axis. The green team generates the fewest opportunities. (h) Increasing the weighting of Cost/WonLead. The green team is the most frugal, but has the most revenue, while the blue team is the most wasteful with not much revenue. (h) STM setup for the animated presentation of these findings.

### 7.1.1 Step 1: explore the PCA view

Let us assume a sales team analyst, Pat, is about to analyze the data. He begins with treating the entire dataset as one cluster and performs PCA – shown in Fig. 11(a). He immediately notices that there are three visually separable clusters representing the three sales teams. These distinct clusters suggest that the three sales teams indeed seem to apply different strategies for possibly different outcomes. Pat clicks the ‘Apply’ button to load the cluster information from the original data. The result (the thumbnail view on the bottom right side of panel (a)) confirms that the three clusters are indeed real clusters.

Next, Pat examines the SE boundary in Fig. 11(a). He notices that there are two groups of closely mapped attributes with strongly printed labels: (1) *Expected ROI* and *Pipeline Revenue*, and (2) *LeadsWon* and *#Leads*. As explained in Section 4.1, this means that the attributes in each of these groups are strongly correlated. Pat finds this view informative and saves it to the STM.

### 7.1.2 Step 2: explore the salient subspaces

Next, Pat wishes to examine the subspaces of each cluster. He performs PCA on all of them and adds them to the STM. Pat then optimizes each subspace such that its focus cluster is best separated from the others. In Fig. 11(b), the view thumbnails outlined in blue, magenta, and green are the subspaces for the correspondingly colored clusters. The neutral view is the subspace for the entire data.

Pat first brings the blue cluster’s subspace back to the SE for closer examination (Fig. 11(c)). He notices that *Cost* has the most prominent label and that the blue cluster varies significantly in this direction – more than the two others. This suggests that there is a wide diversity in the cost incurred by members of the blue sales team.

Next, Pat brings the subspace of the green cluster into the SE (Fig. 11(d)). He notices that in this subspace #Leads and Cost are most widely expressed (i.e. these attributes best distinguish the green sales team from the others). From the plot, Pat learns that the green team, with its cluster being most closely located to the #Leads attribute, seems to generate the most leads, while the blue team generates the fewest. He also confirms the finding from the last view that the blue team seems to incur the highest cost.

Lastly, Pat brings the magenta subspace into the SE (Fig. 11(e)). He confirms some of the findings of the previous plots and also learns that *PlannedRev, Cost* and *ExpectedROI* are the attributes that have the highest variance for this group of data. Finally, he also learns that the magenta sales team is separated from the other two by a combination of *PlannedRev, #Opportunity* and *ExpectedROI.*

#### 7.1.3 Step 3: look for differences in sales strategy

Pat knows that high *Pipeline Revenue* and *Expected ROI* are important targets for any business. He decides that it would be a beneficial undertaking to explore how the company's sales force relates to these two revenue parameters.

He uses the STM to bring the initial PCA view (small panel in Fig. 11(a)) back to the SE. He presses the right mouse button and moves the mouse in the direction of the two revenue parameters. Fig. 11(f) shows the outcome. Note that the font of the two revenue labels gets stronger which means that the corresponding two attributes receive more weight in the viewed 3D subspace. The plot shows that both the green and magenta sales teams generate more revenue than the blue one and that the green team is slighter better than the magenta one. Pat also notices the *#Opportunity* attribute near the top of the plot and that it seems to separate the clusters well. He figures that revenue probably has a lot to do with the generated opportunities and he decides to give this attribute more emphasis.

He uses cluster chasing to emphasize *#Opportunity,* clicking on its label and moving the mouse upwards with the right button depressed. He similarly emphasizes *pipeline revenue* and arrives at a traditional bivariate scatterplot that has *pipeline revenue* along PPA-x and *#opportunity* along PPA-y (see Fig. 11(g)). He observes that while the green and magenta teams vary in the number of opportunities – magenta creates more – both groups have somewhat similar revenue but green has a slight advantage. On the other hand, the blue team also has high *#Opportunity* but its revenue is low.

So why does the blue team lack revenue despite its similar amount of opportunities? Pat knows that sales teams typically spend money to turn won leads into opportunities. He decides to make *Cost per won lead* the new PPA-x axis by selecting it from the attribute list in the control panel (since it is not visible currently). Fig. 11(h) shows the outcome. He quickly notices that the blue team incurs much higher cost than the other teams and that the green team has the highest *pipeline revenue.* In fact, the green team is the most frugal having the narrowest cluster.

Based on these discoveries, Pat concludes that while generating many opportunities sounds like a winning strategy, it is associated with high cost and therefore the generated revenue tends to be low. This is the lesson taught by the blue team. It thus seems better to replicate the green team's strategy – spend little cost on each won lead and, despite gaining fewer opportunities, obtain higher revenue.

#### 7.1.4 Step 3: use the STM for sharing the findings

Pat is excited about his findings and plans a presentation to his group. He notices that the STM is too cluttered and so he uses the "SmallViewSize" slider to reduce the size of the view thumbnails. Then he connects them by simple mouse clicks and builds a path (bottom panel in Fig. 11(i)). Clicking the 'Next' button, all his findings can now be displayed sequentially, in an animated fashion.

#### 7.1.5 Conclusions from this use case

We believe that this example convincingly demonstrates how our SE interface enables users to playfully arrive at different multivariate scatterplot projections, quickly respond to new explorations ideas on a whim, make casual observations in the process, and just as easily return back to a traditional bivariate scatterplot visualization. The interested reader may watch the video to see the complete process.

### 7.2 Use Scenario #2: Visual Item Discovery & Selection

Selecting the best college, given the many personal constraints and preferences one might have, is arguably one of the most difficult choices a person will make in life. It involves the task of discovering the set of schools that best meet one's personal requirements, comparing them by weighing their trade-offs, and then selecting the college that fits best. Here we use the mixed dataset initially created by Nam and Mueller [31]. It has multi-faceted data on 50 of the top US colleges, enabling the college-seeking student to look at schools not only through the lens of academics but also through the lens of social life and the general environment the school resides in. Academic ranking and tuition information were extracted from a leading source of such information – the US News & World Report [49]. The College Prowler website [48], on the other hand, ranks colleges on a multitude of social and environmental factors. We picked 8 of the 20 the site offers: athletics, campus housing, local atmosphere, nightlife, safety, transportation, academic environment, and weather. Each score is available letter-graded ranging from A+ to D-. We mapped these equidistantly to values in the range 0 to 1.

The College Prowler website allows users to navigate the space of college attributes by filtering, using slider bars and menu selections for each parameter to narrow down the search. This can be rather tedious and it also makes it difficult to recognize tradeoffs. We believe that our SE provides a more playful and targeted experience, while the STM is a better platform to save any intermediate findings.

In the following, we shall follow 17-year old Tina who is just about to finish high school and see how she uses our subspace voyager to find the university she feels best about.

#### 7.2.1 Checking out the relationships of attributes

Tina starts out with a view onto the dataset as a single cluster using the primary PC axes as a basis (Fig. 12(a)). As mentioned in Section 4.1, in such a view the dimension vectors of strongly positively correlated attributes tend to coincide and as a result, their labels map to similar locations along the trackball boundary. Conversely, negatively correlated attributes will map to opposite sides of the trackball boundary. The only condition for both is that their projection into the PC-axes basis is sufficiently significant, which is visually expressed in our system by a large and heavy label font. In the initial view of Fig. 12(a) Tina observes two sets of positively correlated attributes: (1) *Academics* and *Tuition*, and (2) *LocalAtmosphere*, *NightLife,* and *Transportation*. She also observes a few negatively correlated attributes, among them: (1) *Academic* with *Weather* and *Athletics,* and (2) *LocalAtomosphere* and *NightLife,* with *Safety*. From these constellations, Tina quickly recognizes that top academic universities tend to charge higher tuition, but at the same time, their athletic teams are not necessarily among the best. She also learns that universities built in nice town or city areas usually have better nightlife and transportation systems, but they also tend to be less safe. All this is good to know before engaging in the actual selection process described next.

#### 7.2.2 Finding the set of schools that fit the best

Tina does not come from a wealthy family and so her immediate focus is tuition cost. Her first step is, therefore, to select *Tuition* and move the mouse towards that label (to the left). Next, she wants to see which of the schools have a good academic ranking. She selects *USNewsScore* and moves the mouse downward to maximize the spread. This leaves her with the axis-aligned scatterplot shown in Fig. 12(b). In this plot, all points on the lower right side are the universities with high rankings but low tuition – these are the ones Tina is interested in the most. She colors them in magenta and asks the system to label them – in this case with the university names.

Tina likes the outdoors a lot which requires the weather to be generally good. So she adds *Weather* as another requirement to

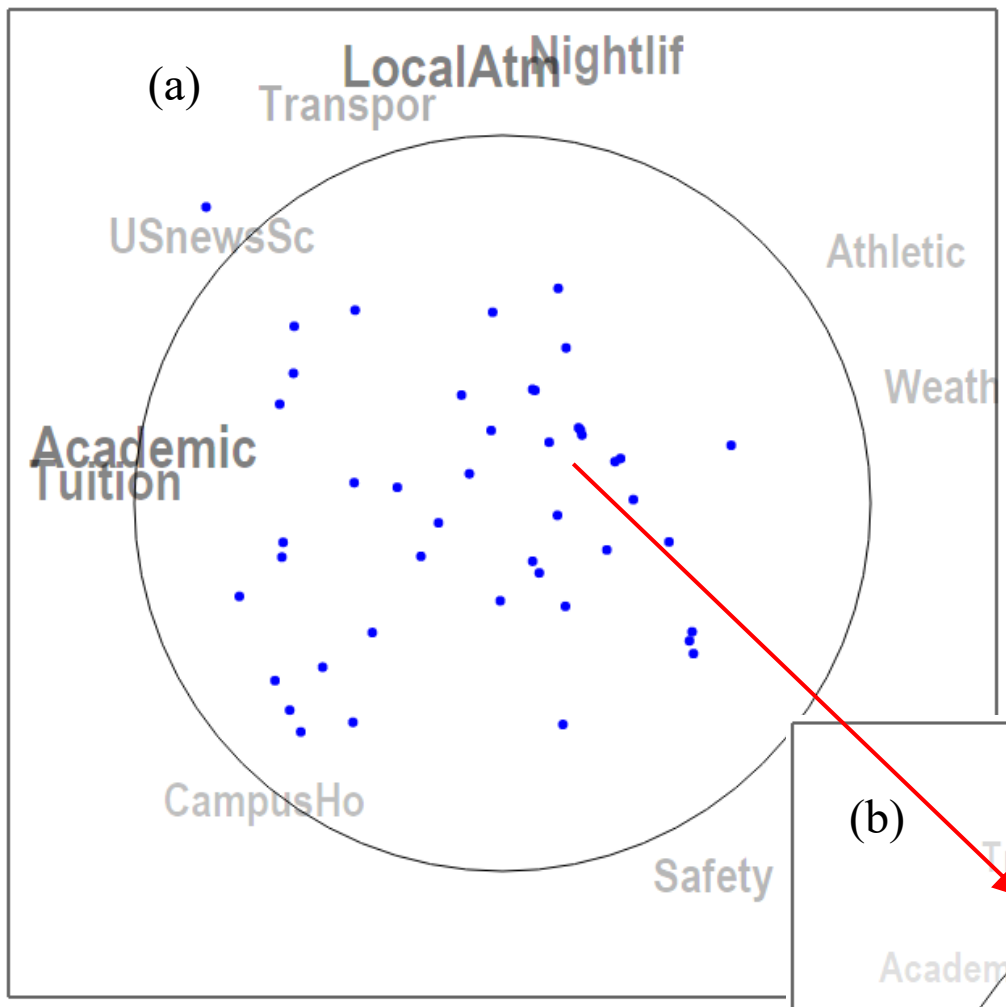

(a). The PCA view of the entire dataset. It reveals a strong positive correlation between *Academic* and *Tuition*, as well as between *LocalAtmosphere, NightLife,* and *Transportation*.

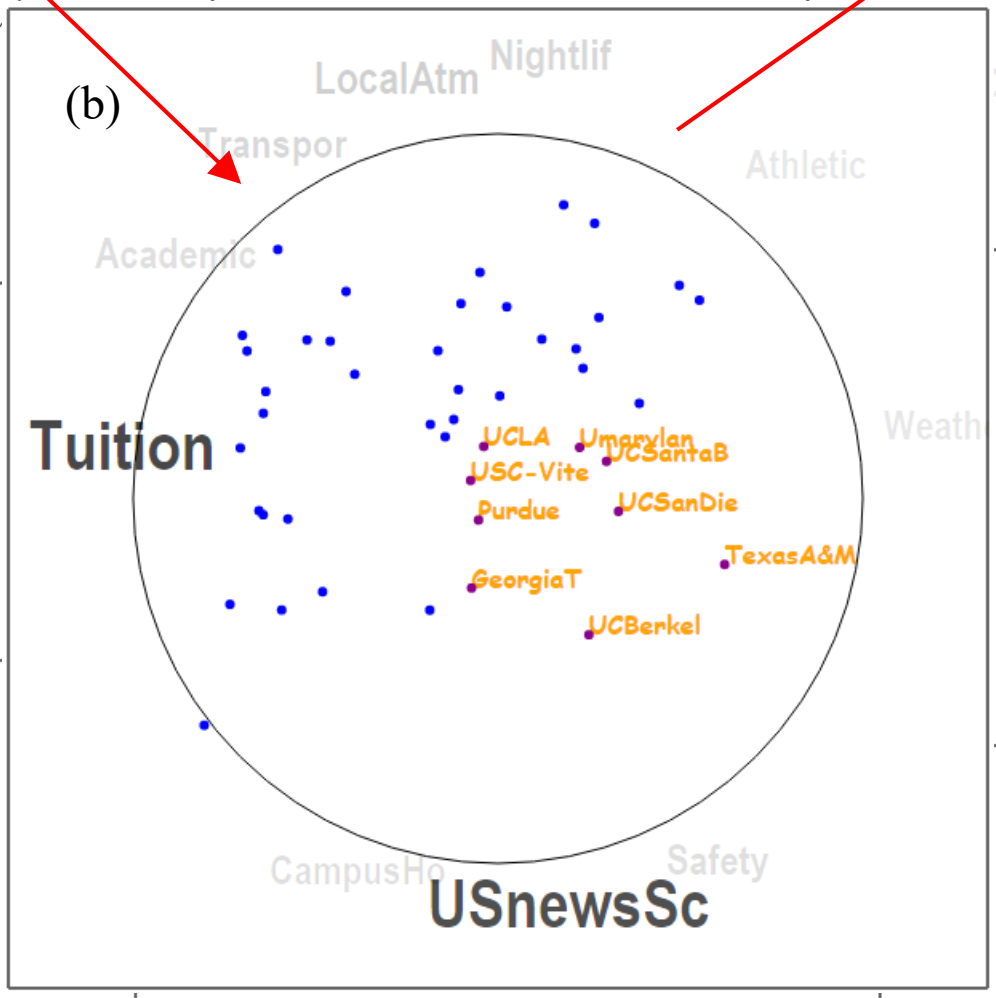

(b). Make *Tuition* the PPA-x axis and *USnewsScore* the PPA-y axis. The magenta-colored, labeled points are the schools with high *US News Score* and low *Tuition.*

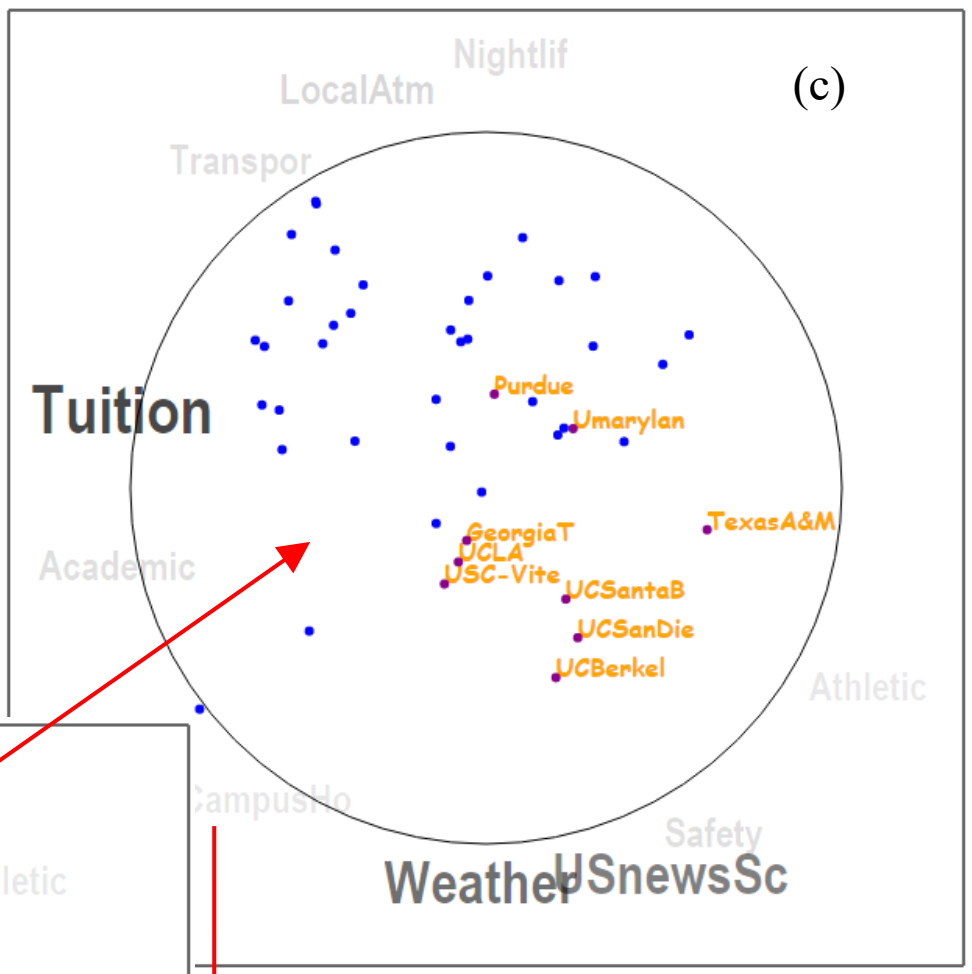

(c). Increase *Weather* along the PPA-y axis. Purdue moves up revealing its bad weather, TexaxA&M and UMaryland do not move much but are low in both scores.

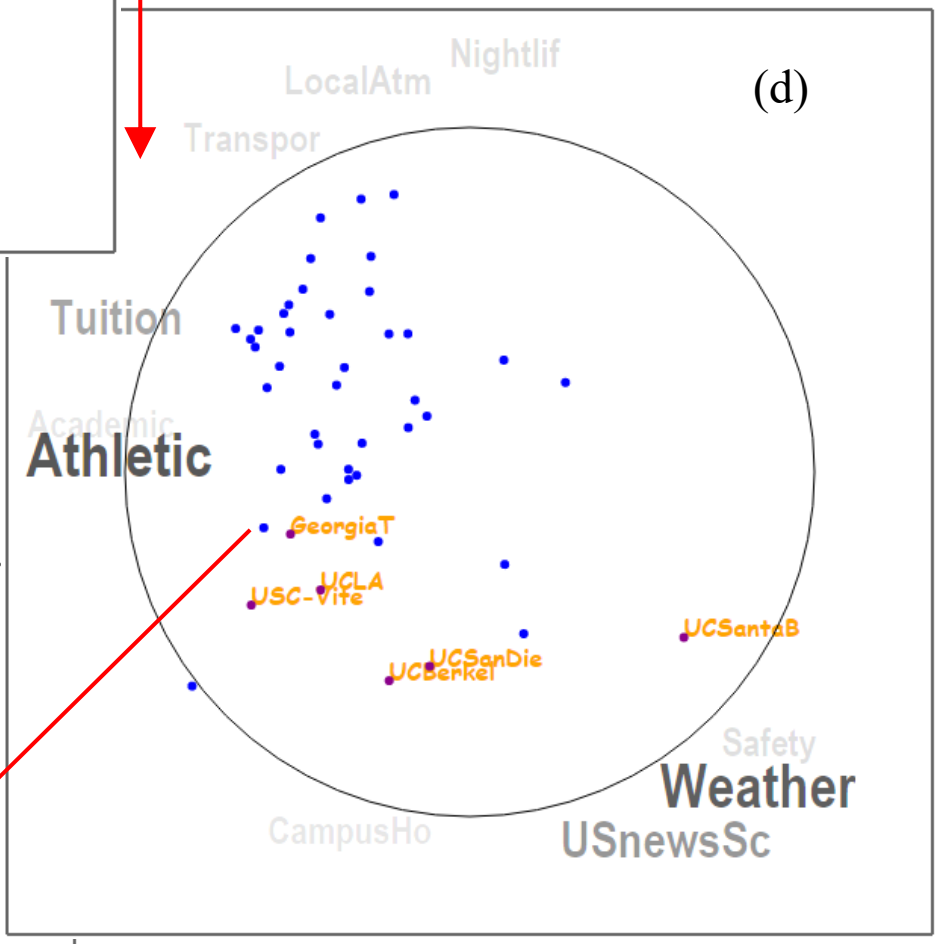

(d). Increase *Athletic* along the PPA-x axis. UCSantaBabara has a very low athletic score.

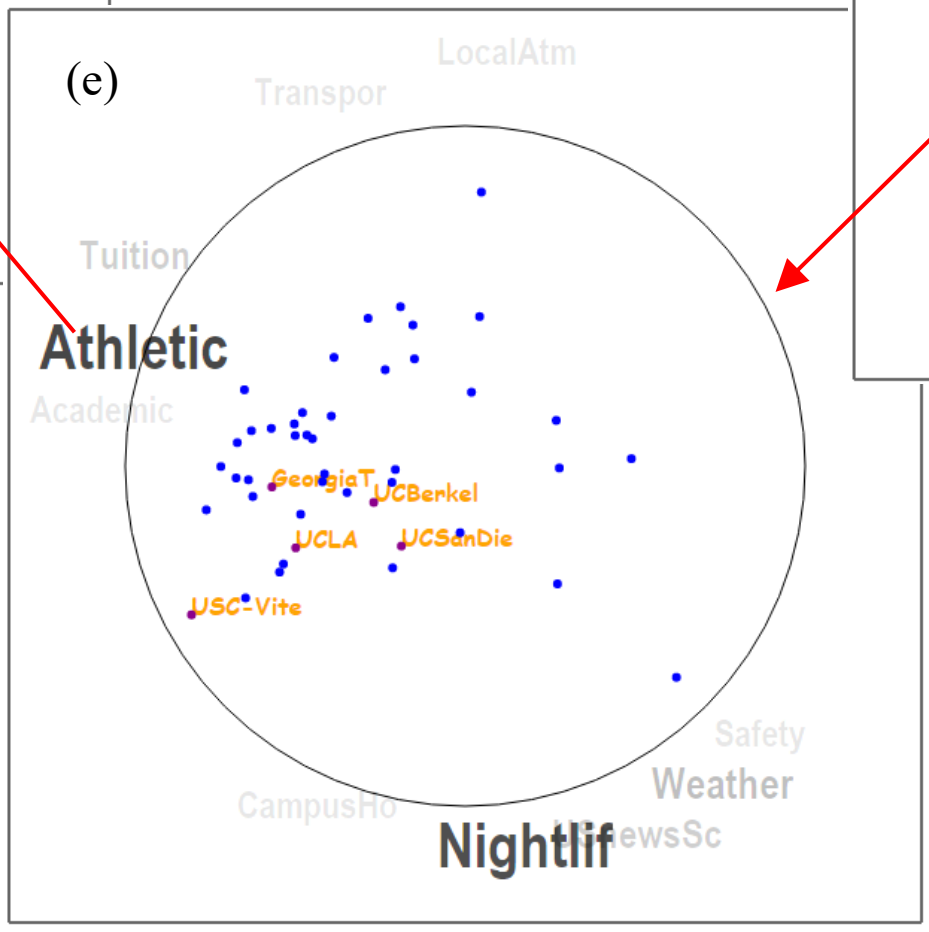

(e). Increase *Nightlife* along the PPA-y axis. USCViterbi has the best overall score, followed by UCLA, Georgia Tech, UCSanDiego and UC Berkeley.

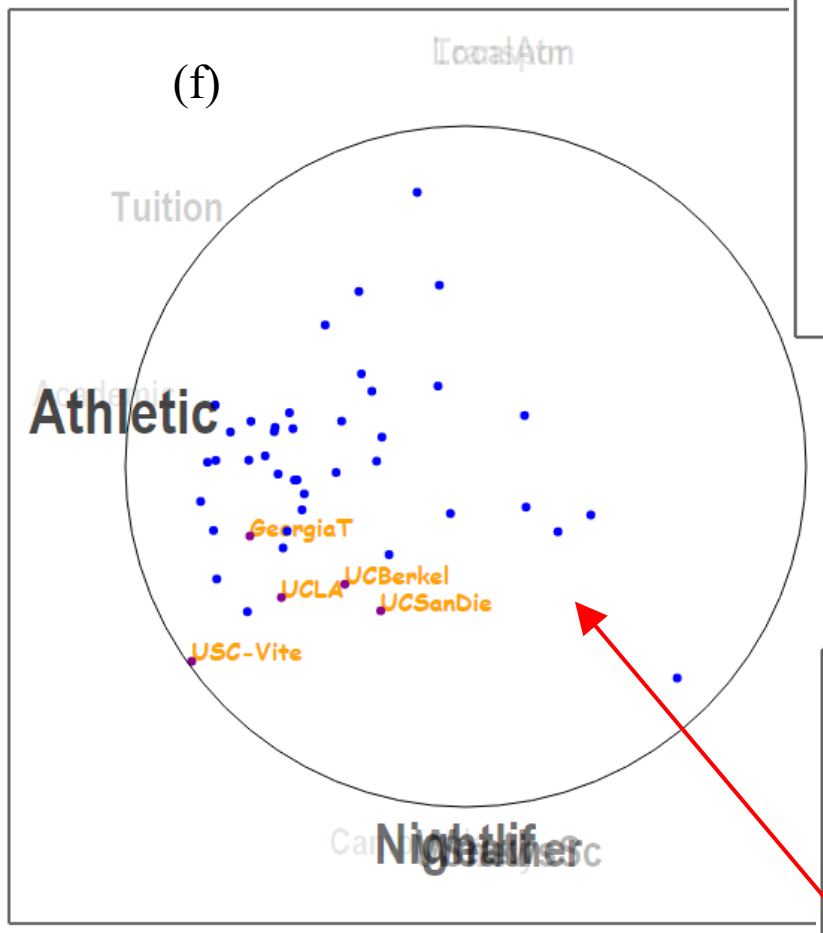

(f). The final setup. The dimension projection reveals four dimension groups.

Fig. 12. Finding the best college in the college dataset.

Table 1
Rankings of the final five candidates

| College | Acad. | Athletics | House | Atmos. | Night Life | Safety | Trans. | Weather | US News | Tuition |
|---|---|---|---|---|---|---|---|---|---|---|
| **UCLA** | 10 | 10 | 5 | 12 | 10 | 9 | 4 | 11 | 69 | 22428 |
| **USC-Viterbi** | 10 | 2 | 8 | 11 | 12 | 7 | 8 | 11 | 77 | 22734 |
| **Georgia Tech** | 10 | 11 | 5 | 10 | 9 | 7 | 7 | 8 | 86 | 22188 |
| **UC Berkeley** | 10 | 9 | 8 | 9 | 8 | 7 | 10 | 11 | 89 | 14998 |
| **UC San Diego** | 9 | 8 | 6 | 11 | 9 | 11 | 6 | 12 | 72 | 14694 |

*USNewsScore* by selecting the *Weather* label and moving the mouse in the downward direction (Fig. 12(c)). This enables her to appreciate any tradeoffs that may exist in these two variables.

After making this choice, she sees 'Purdue' moving away significantly. This means that even though its *USNewsScore* is quite good, it is not good enough to make up for the unfavorable weather. Likewise, 'GeorgiaTech' also moves away but not as much and so Tina keeps it marked and labeled. 'UMaryland' and 'TexasA&M' did not move too much either, but both of their scores are not high from the onset. It means that none of these two schools does well enough in either *UsnewScore* or *Weather* to make up for the moderate performance in the other attribute. Tina removes these two schools as well, repainting them to neutral blue.

Tina enjoys the excitement of college team sports. She is also quite athletic and she thinks she might be able to secure an athletic scholarship to pay for her tuition. So she puts the *Athletics* attribute near the tuition using the aforementioned mouse interactions (Fig. 12(d)). She notices that 'UCSantaBarbara' has a rather poor athletic score and henceforth she eliminates that school. On the other hand, 'USC-Viterbi' has the highest athletic score, followed by 'GeorgiaTech', 'UCLA', 'UCBerkely' and 'UCSanDiego'. She keeps all them magenta colored and labeled.

Of course, Tina wants to have some fun in college. She focuses on *NightLife* and moves it to the bottom (Fig. 12(e)). 'USCViterbi' moves down confirming that it has a good nightlife. 'UCBerkeley' and 'UCSanDiego' move up, indicating that they may have good weather but the nightlife is limited. Conversely, 'GeorgiaTech' and 'UCLA' stay put – they are more balanced in those two factors.

#### 7.2.3 The decision – selecting the #1 school

Looking at the plot shown in Fig. 12(e) Tina sees that 'USC-Viterbi' might be the best candidate. It is somewhat in the middle between *Athletics* and *Nightlife* and it is closest to the circle boundary indicating that it has the highest values there. Yet, 'GeorgiaTech' and 'UCLA' are both not far behind and could be close contenders. In order to gain an overall impression, Tina puts all attributes of interest into one view. She tilts the trackball and creates the configuration shown in Fig. 12(f). Four dimension groups emerge (1) *Athletic* and faintly *Academic*, (2) *Tuition*, (3) *LocalAtmosphere* and *Transportation* (both with small weighting), (4) *NightLife*, *Weather* and faintly *USNewsScore*, *Safety,* and *CampusHousing*. Among all those groups of factors, Tina values *Athletic* and *Academic* the most, and so she chooses 'GeorgiaTech' as her #1 top choice school to apply for.

#### 7.2.4 Comparison with TripAdvisor$^{ND}$

We purposely conducted a similar selection task than Nam and Mueller in [31], and a partial goal of this use case was to compare the two systems. We obtained rather similar, almost identical results than these authors, except that their final candidate list did not contain 'UCLA'. We compared UCLA's scores with that of the other candidates (see Table 1) and found that except for a lower rating in transportation and a slightly lower rating in *USNewsScore,* it is not worse in other aspects and hence should be included in the final candidate set. Especially in the final dimension group *Academic* and *Athletic*, it has a better combination of values than the other schools in the set, except for 'Georgia Tech'.

We think that the omission of 'UCLA' occurs because TripAdvisor$^{ND}$'s motion trail makes it sometimes difficult to precisely follow each point. But the motion trail is needed there to convey the dynamic movement. Conversely, with our trackball, motion trails are not required since the perception of the motion is much more tightly linked to the interaction that is causing it – both occur in the same interface.

Another advantage of our new system is that users no longer need to take their eyes off the visualization while they are interacting to change the view. TripAdvisor$^{ND}$'s touchpad required this. It also required that two points are moved separately – the one due to the PPA-x and the one due to the PPA-y vector. With our trackball interface, users can express these goals much more directly. In fact, they do not even need to be aware of the existence of these axes and vectors which we believe makes our interface much more appealing to general users.

### 7.3 Use Scenario #3 – Visual Cluster Refinement

Often high-D data are derived from feature analysis where the features themselves are not overly meaningful in isolation. Rather, it is the synthesis of all features that allow users to describe a grouping of the data points. In this process, the feature-based clustering provides the organization in which the boundaries of the individual groups can be delineated. We now demonstrate how our system can be used to allow humans to assist in creating and refining these kinds of groupings in data, using visualization as a gateway. We call this process *visual cluster refinement*.

We have selected an image classification tasks for this use case. The CLEF Cross-Language Image Retrieval Track (ImageCLEF) [50], launched in 2003, is an evaluation forum for the cross-language annotation and retrieval of images. It provides a language independent platform where visual information retrieval systems can be evaluated for analysis, classification and retrieval tasks. The ImageCLEF data [15][51] entails three sets of images – training, testing, and development. Each set uses different feature descriptors to describe the images, such as SIFT, color histogram, and GIST. We use the GETLF feature vector from the development set of ImageCLEF 2013 [52], which is a 256-dimensional histogram based feature. For the exact information on how to generate these descriptors, the reader is referred to [15].

In the following, we employ our Subspace Voyager as a medium to bring users into the loop of assessing and assisting the process of top-down clustering of this dataset. Since the cognitive processes driving image recognition and assessment are still much better developed in humans (as opposed to machines) a visual interface that allows humans to participate in this task can be very valuable.

We begin by setting the initial number of clusters to a value of 2 and run k-means clustering with the structure-based distance metric [25] on the collection of points. Fig 13 (a)(b) shows the two subspaces in which the two clusters reside. Since the attribute labels on the trackball boundary are rather cryptic, a visual assessment of cluster quality is difficult and even more so is their interactive refinement. This can only be done by visualizing the semantics of the data points themselves – in this case, the underlying images.

To facilitate this, we examine the two clusters separately inside their own subspaces by turning off the other cluster. Similar to Liu et al. [27] we randomly select a subset of the data points in each cluster and display the corresponding images next to them (Fig. 13(c)(d)). We notice that the images in the magenta cluster (Fig. 13(c)) are overall more saturated than those in the blue cluster (Fig. 13(d)). For the magenta cluster in Fig. 13(c) a clear change from yellow to black can be observed along the PPA-y axis, with yellow sunsets, yellow flames, and yellow-leaved trees on the bottom of the distribution, mixed with increasingly more black towards the top. For the blue cluster in Fig. 13(d), the images on the bottom left are a paler green mixed with gray while those on the top are mostly blue toned. The images on the right half have a background that is mostly white, with the main objects being low saturated. The differences between the two clusters in Fig. 13 (c, d) are obvious and this confirms that k-means can separate the dataset well at this level of the hierarchy.

We now continue this process and further partition the clusters using k-means. We choose the blue cluster (Fig. 13(d)) since its diversity is much greater than that of the magenta cluster. As there seem to be three main categories of images we pick k=3. Figs. 13(e-g) show the results. We observe that the classification has become more specific. The sub-cluster in Fig. 13(e) has the blue images, fading out towards the bottom left. The sub-cluster of Fig. 13(f) has the green images, again fading out towards the bottom left. Finally, the sub-cluster in Fig. 13(g) has the saturated images or the images with white background. While this 2-level tree could already suffice for some applications, one more level of sub-clustering might separate the fully saturated images from the less saturated ones in each of the clusters obtained so far. Here the visualizations can help to determine where and whether such a step is required or desirable.

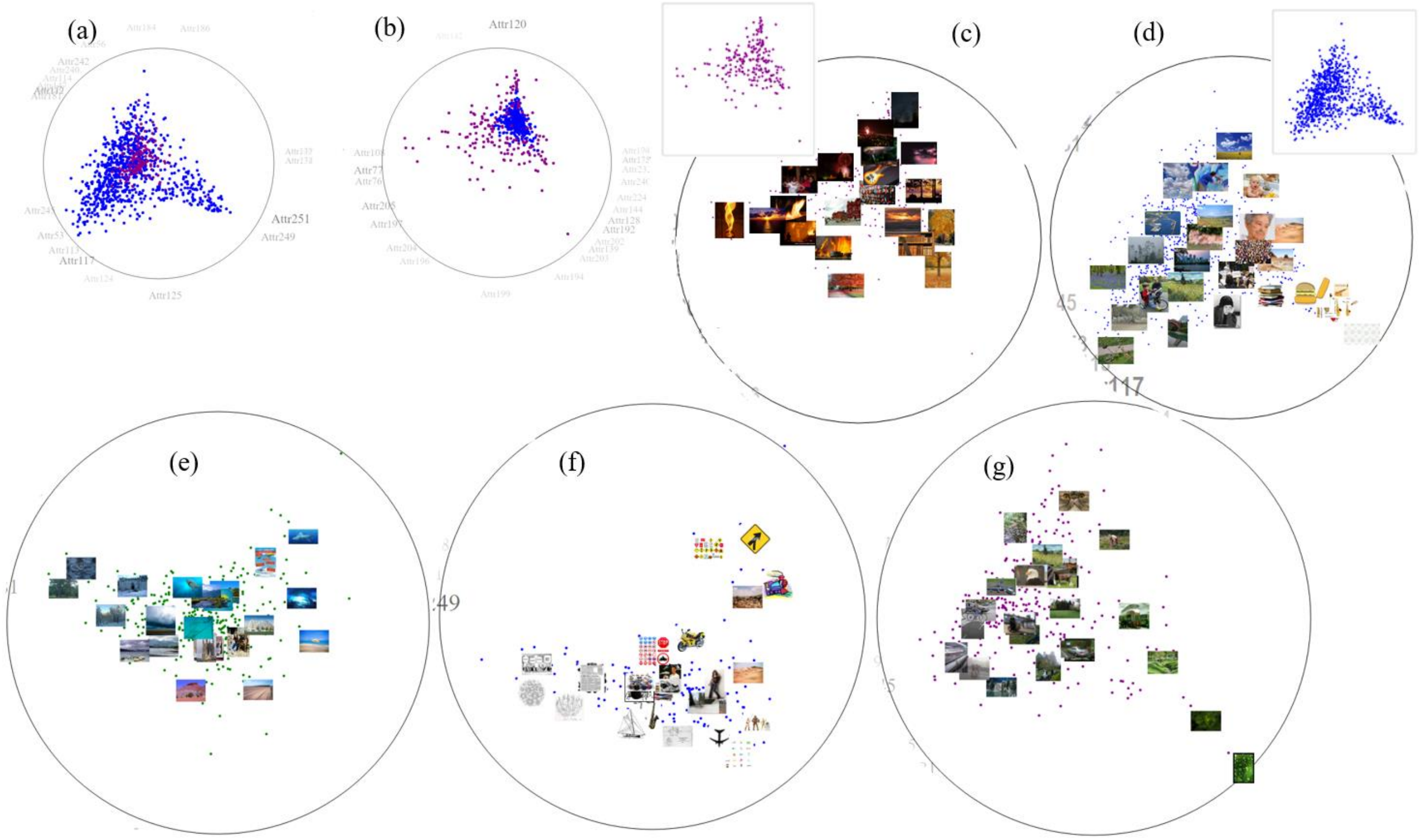

Fig.13. Exploring the ImageCLEF dataset. (a) The subspace of the blue cluster. (b) The subspace of the dark magenta cluster. (c) The images in the dark magenta cluster. (d) The images in the blue cluster. (e-g) The three sub-clusters of the cluster in (d).

## 8. USER STUDY

To evaluate the features of our interface, we conducted a user study with 10 graduate students – 3 females and 7 males of diverse cultural background (4 North Americans and 6 Asians). None had experience in visual analytics. We sought to test whether the participants could fulfill certain data analysis tasks using our system.

### 8.1 Setup

We invited all participants to sit down with us individually. We first showed them a 3-minute intro video which covered all basic interactions the Subspace Voyager supports – the trackball interactions, saving views to the STM and bringing them back to the SE, and traversing between STM views. We used the Iris dataset [53] as a walk-through example in this video.

After the video, each participant could ask questions to resolve any doubts. This was the only time we entertained questions. Three participants were unclear on how to interpret the dimension labels along the trackball -- a brief explanation resolved these doubts. Next, each participant was asked to perform three tasks. With their consent, we filmed the computer screen to record their operations. We asked them to speak out their thoughts during the entire time, which we also recorded. We used both recordings in our study.

### 8.2 Tasks

Our tasks were designed to measure three levels of understanding [19] gained from the visual interactions – shape, organization, and relations. For the first task, we used a somewhat contrived 3D data set: a hollow tube with a stick in the middle that was not aligned with any of the data axes. We asked the participants to describe the shape of the data, first using the SPLOM (Fig. 14(a)) and then using the Subspace Voyager (Fig. 14(b)).

The second task measured visual understanding. We did not inform the participants about the nature of the dataset. We initialized the STM with two scatterplots of the sales force data (Fig. 15(a-b)) and asked the participants to describe what they saw in these two plots. Then we encouraged them to transition between the two scatterplots and again asked them to describe their impressions.

For the data understanding task, we also used the sales force dataset but now we first showed the participants a 1-minute video that introduced the attributes of the data. The initial trackball configuration is shown in Fig. 16(a). In this view, the clusters for the three teams overlap. The participants were told that there were three sales teams, who used different sales strategies and that the task was to determine which of the strategies was best to reach high revenue.

### 8.3 Result

The following results also include a comparative study with the TripAdvisor$^{ND}$ interface, using the same data and participants.

#### 8.3.1 Task 1 – Shape understanding

Not a single participant found the hidden stick in the SPLOM. This was to be expected since this structural feature can only be observed from a non-axis aligned angle. Eight participants asked for pen and paper to reconstruct the distribution of the data but none got it completely right. Descriptions were 'tilted cylinder' or 'oval prism'.

On the other hand, using our trackball's rotation functionality, all participants managed to find the hidden stick. They spent 48 seconds on average for this task. Some of the descriptions given were 'pipe with something in the middle', 'cylinder with some coaxial cable', or 'two concentric cylinders'. Fig. 14(c) and (d) shows two typical views the participants generated and which helped them draw their conclusions. This high success rate demonstrates that our trackball is able to help users understand the structure of data.

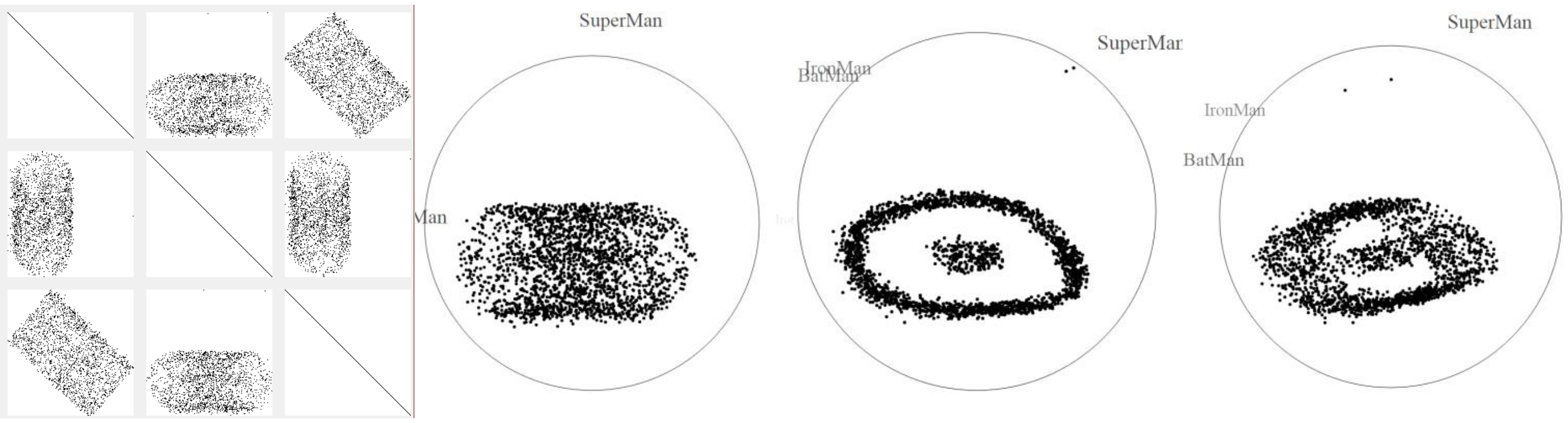

Fig. 14. User study 1: shape understanding. We asked the participants to use both (a) SPLOM and (b-d) our system to examine the shape of the data. (b) The initial view in the SE. (c, d) Two typical views the participants generated to help them draw their conclusions.

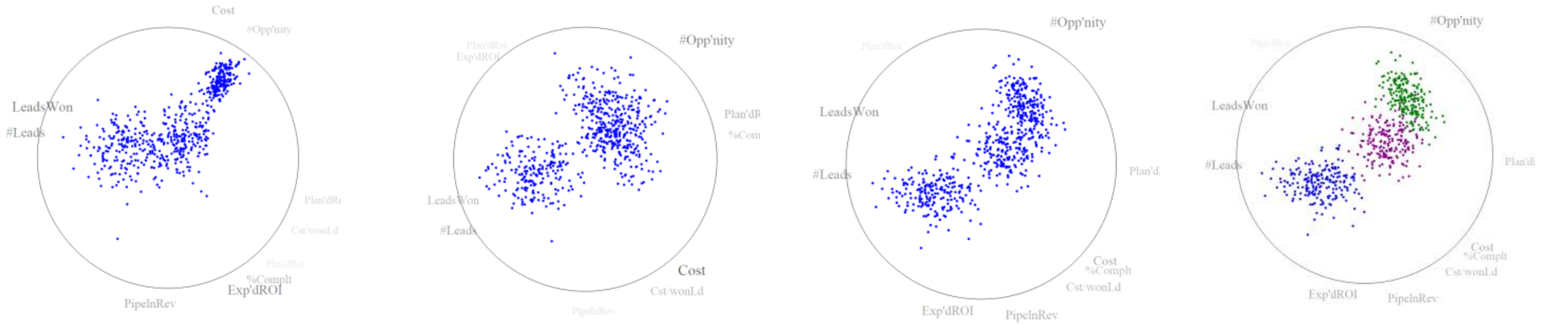

Fig. 15. User study 2: visual understanding. We showed the participants an STM composed of two views (a) and (b). We then asked them to traverse between these two views and describe what they saw along the path. This path included view (c) which was the most revealing – the motion parallax clarified that there were indeed three clusters. (d) The three clusters marked in different colors as a confirmation.

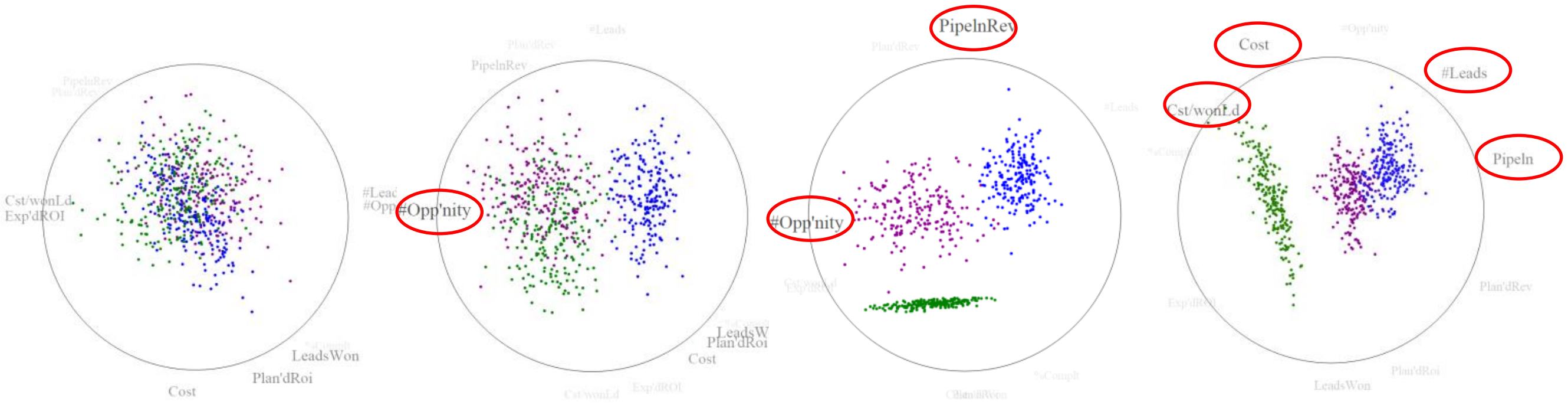

Fig. 16. User study 3: data understanding. (a-d) Various views our participants generated. See Section 8.3.3 for a narration of these plots.

### 8.3.2 Task 2 – Visual understanding

Starting out with the two view thumbnails (shown expanded in Fig. 15(a-b)) eight of the participants stated that there were two clusters. On the other hand, two of the participants suspected that there might be a third cluster, mostly based on the view in Fig. 15(a). Next, all used the 'TraverseBtw' slider to go from one view to the other (nine users went from the top view to the bottom one while one went the other way). Fig. 15(c) has the most revealing view they generated (colored in Fig. 15(d)). Everyone spotted the third cluster while traveling. The average time for completing the task was 83s. Most of this time was spent on describing the observations while only 10–20s was spent on traversing between the two subspaces.

The transition interface seemed very effective in helping the participants understand the high-D structures. Some of the comments were 'The bigger cluster separates into two, one of them remains a separate cluster, while the other one merges with the smaller cluster' and 'The upper left cluster seems to be moving forward. Another cluster is moving upward, and the third one is moving downward'. One person saved a couple of views in the STM and later mentioned that 'if I look at those still frames, I probably still cannot tell that there are three clusters, it's really the motion by which you can tell.' This comment nicely verbalizes the power of motion parallax. It makes it easier to spot patterns than relying purely on still frames.

### 8.3.3 Task 3 – Data understanding

For this task, we colored the points according to sales team. We observed that the participants often used the 'chase cluster' functionality but as a part of different exploration strategies. One of these strategies consisted of making individual data dimensions the dominant factor on either the PPA-x or the PPA-y axis and then observing the distribution of the three sales teams along these attributes. Fig. 16(b) shows a view generated by a participant who wanted to examine the influence of *#opportunities*. Later on, when using the STM to return to this view to present her findings, she stated that 'the blue team has the lowest number of opportunities'.

Another strategy was to keep *PipelineRevenue* as either the PPA-x or PPA-y axis and assign another attribute to the other axis. By doing this, these participants managed to create traditional axis-aligned scatterplots where they could examine the relationships between two variables. One typical view of this method is shown in Fig. 16(c). Here the participant concluded: 'The blue and magenta teams all have high revenue. The magenta and green teams all have a high number of opportunities.'

Yet another analysis strategy was to try to come up with certain views where some attributes, or all of them, were well expressed. A typical view for this strategy is shown in Fig. 16(d). Based on this view, the participant generating it described his findings as 'The blue and magenta teams have higher revenues than the green one. They both generate more leads and have a lower cost per won lead. Those two factors seem to be important.'

All participants used the STM to save important findings. When they reported these finding to us, six participants simply dragged and dropped their saved views back to the trackball, while four participants used the STM to traverse between these key frames when telling us the story because they 'liked the animation'. When presenting their findings, all participants arrived at similar conclusions, viz. 'The blue and the magenta teams have the highest revenue', 'The blue team has a high number of leads and this might lead to their high revenue' and 'Cost per won lead needs to be low'.

The participants spent on average 17 minutes and 48 seconds on this task. There was a wide time spread. One of the participants spent almost 33 minutes because he was 'having fun' and just wanted to explore the dataset more. We believe this proves that our system is very playful and nicely engages the users into the data analytics.

#### 8.3.3 Further finds from the voice recordings

Analyzing the recordings produced further interesting comments. One participant said our system was 'very intuitive' and 'helped me not only understand the distribution of the data on one specific dimension but multiple dimensions at the same time'. Other participants stated that our system made 'exploring data fun', that the trail map made 'switching back and forth very fast', and that this was the 'first time seeing data exploration could be done in this way'.

#### 8.3.4 Comparative study with TripAdvisor$^{ND}$

We also conducted a comparative study with the TripAdvisor$^{ND}$ interface to test if our system could outperform its ancestor. Since TripAdvisor$^{ND}$ does not have a trail map for the navigation between different subspaces, we only repeated the first and the third tasks. We engaged the same participants as in the Subspace Voyager study and we used the same data. There were, however, six months between the two studies. This greatly diminished learning effects with respect to the data. Our goal was primarily to learn about pros and cons of the two systems, and only get a rough estimate of the time needed to accomplish views comparable to those obtained before.

For the first task, the participants expressed that the subspace voyager was more 'direct'. They thought that the pad navigation interface in TripAdvisor$^{ND}$ 'made the control of the shape difficult. We observed that the participants were mostly moving the navigation points 'arbitrarily' until they found out what was going on. One participant mentioned that because of the 'extra layer of the interface, how to control the points is less obvious and less intuitive'. We asked him what he meant by 'extra layer'. He replied 'it's the separate navigation interface and also the switching between x and y control points'. On average, the participants spent 91 seconds to finish this task using TripAdvisor$^{ND}$. This is about twice the time they needed with the Subspace Voyager.

For the third task, the one participant who chose to make individual data dimensions the dominant factor on either the PPA-x or PPA-y axis preferred the navigation polygon in TripAdvisor$^{ND}$. He thought moving the red and blue dots onto any two dimensions was very straightforward in TripAdvisor$^{ND}$ while 'it required a bit more adjustment to single out the two dimensions I want' in Subspace Voyager. In contrast, the other participants who tried to come up with certain views where some attributes, or all of them, were well expressed all preferred Subspace Voyager. They all agreed that it was much easier to come up with a meaningful multivariate projection using the new system. One participant said that for TripAdvisor$^{ND}$, 'to get the superposition of dimensions, I had to move the vertices of the dimensions I am interested in next to each other in the polygon and it's tedious' while 'In the Subspace Voyager, moving dimensions to desired locations and controlling their weights was very straightforward'.

We believe that these (mostly qualitative) findings and assessments make a conclusive argument for the Subspace Voyager.

## 9. Conclusions

We demonstrated a system for high-D data exploration in form of scatterplot projections that decomposes the high-D data space into a continuum of generalized 3D subspaces. Using 3D space as the immediate visual context affords a natural user interface well suited for mainstream users. The interactive tools we designed do not require users to ever think of data in their native high-D context. Rather, users fluidly transition from one generalized 3D subspace to the next in a goal-directed manner, emphasizing and de-emphasizing the weights of the various attributes on the fly during the visual interactions. Key elements of our system are an augmented trackball with a peripheral weight-adaptive attribute label display, a metric-driven view and subspace optimizer, and a map that allows users to organize the scatterplots of key findings and transition between them. We also provided several measures that help scalability for both attributes and data items. Users can control the number of attribute labels shown and they can hide data points temporality to improve the visibility of the points currently deemed relevant. Several user studies confirmed that our system supports both visual and data understanding.

In future work, we would like to add depth information when displaying the data points in the trackball. This could be achieved by introducing depth of field effects and fog and converting the point clouds into shaded 3D shapes with drop and self-shadows, combined with occlusion and semi-transparent effects. All of these could allow users to better appreciate the structure of the data and discover patterns that would otherwise be hard to notice in a conventional scatter plot display. The end goal of all of these efforts is to create intuitive displays, which invite discussion of personal findings with colleagues in business planning and policy-making scenarios, etc. We also plan to refine our system via percept-oriented studies [12].

## 10. Acknowledgements

This research was partially supported by NSF grants IIS 1527200 and IIS 1117132, as well as the MSIP (Ministry of Science, ICT and Future Planning), Korea, under the "ITCCP Program" directed by NIPA. We thank Eric Papenhausen for proofreading the manuscript.

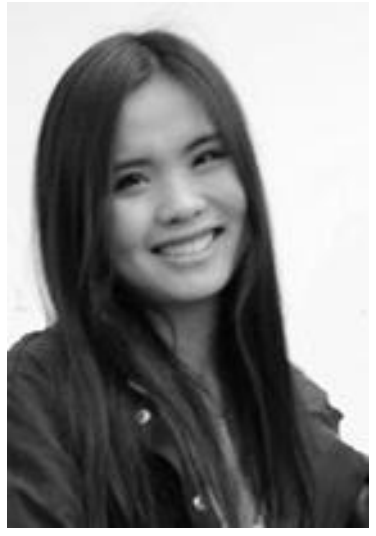

**Bing Wang** received her PhD from the Computer Science Department at Stony Brook University. Her research focus is high dimensional data visualization and visual analytics.

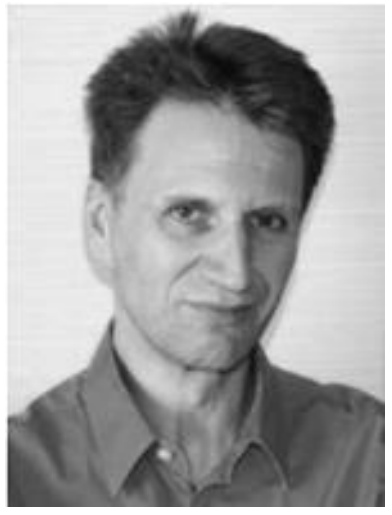

**Klaus Mueller** received a PhD in computer science from The Ohio State University and is currently a professor of computer science at Stony Brook University. His research interests are visual analytics, HCI, and medical imaging. He won the NSF CAREER award in 2001 and the SUNY Chancellor Award for Excellence in Scholarship and Creative Activity in 2011. For more info see http://www.cs.sunysb.edu/~mueller/